\documentclass[11pt]{article}
\usepackage{amsmath,amssymb,amsfonts}
\usepackage[normalem]{ulem}
\usepackage[nosort]{cite}
\usepackage{graphicx,color}
\usepackage[colorlinks=true, linkcolor=blue, citecolor=blue, linktoc=all]{hyperref}
\usepackage[usenames,dvipsnames]{xcolor}
\usepackage{sfmath}
\usepackage{tikz-cd}
\usepackage{pict2e}

\makeatletter
\DeclareRobustCommand{\loplus}{\mathbin{\mathpalette\dog@lsemi{+}}}
\DeclareRobustCommand{\lotimes}{\mathbin{\mathpalette\dog@lsemi{\times}}}
\DeclareRobustCommand{\roplus}{\mathbin{\mathpalette\dog@rsemi{+}}}
\DeclareRobustCommand{\rotimes}{\mathbin{\mathpalette\dog@rsemi{\times}}}

\newcommand{\dog@rsemi}[2]{\dog@semi{#1}{#2}{-90,90}}
\newcommand{\dog@lsemi}[2]{\dog@semi{#1}{#2}{270,90}}
\newcommand{\dog@semi}[3]{%
  \begingroup
  \sbox\z@{$\m@th#1#2$}%
  \setlength{\unitlength}{\dimexpr\ht\z@+\dp\z@\relax}%
  \makebox[\wd\z@]{\raisebox{-\dp\z@}{%
    \begin{picture}(1,1)
    \linethickness{\variable@rule{#1}}
    \roundcap
    \put(0.5,0.5){\makebox(0,0){\raisebox{\dp\z@}{$\m@th#1#2$}}}
    \put(0.5,0.5){\arc[#3]{0.5}}
    \end{picture}%
  }}%
  \endgroup
}
\newcommand{\variable@rule}[1]{%
  \fontdimen8  
  \ifx#1\displaystyle\textfont3\else
    \ifx#1\textstyle\textfont3\else
      \ifx#1\scriptstyle\scriptfont3\else
        \scriptscriptfont3\relax
  \fi\fi\fi
}
\usepackage[makeroom]{cancel}
\usepackage{slashed,youngtab}

\newcommand{\mX}{\mathfrak{X}}
\newcommand{\mY}{\mathfrak{Y}}

\newcommand{\mh}{\mathfrak{h}}
\newcommand{\mw}{\mathfrak{w}}

\DeclareRobustCommand{\loplus}{\mathbin{\mathpalette\dog@lsemi{+}}}

\newcommand{\hatd}{\hat{d}}
\newcommand{\un}[1]{\underline{#1}}

\newcommand{\chkM}{{\color{red} \,\checkmark\kern-5pt{}_{M}}}

\newcommand{\be}{\begin{equation}}
\newcommand{\ee}{\end{equation}}
\newcommand{\beq}{\begin{eqnarray}}
\newcommand{\eeq}{\end{eqnarray}}
\newcommand{\bea}{\begin{eqnarray}}
\newcommand{\eea}{\end{eqnarray}}
\newcommand{\beqn}{\begin{eqnarray}}
\newcommand{\eeqn}{\end{eqnarray}}

\def\pa{\partial}

\newcommand{\hlt}[1]{{\color{WildStrawberry}{\em #1}}\index{#1}}
\newcommand{\dM}{\mathfrak{diff}(M)}
\newcommand{\dS}{\mathfrak{diff}(S)}

\newcommand{\RR}{\mathbb{R}}

\newcommand{\aff}{{{\cal B}}}

\newcommand{\thistitle}{Universal Corner Symmetry and the Orbit Method for Gravity
}
\newcommand{\ulb}[1]{
	\centerline{
		\begin{minipage}[c]{0.7\textwidth}
			\begin{center}~
			${}^{#1}$ Physique Math\'ematique des Interactions Fondamentales \& International Solvay Institutes, Universit\'e Libre de Bruxelles, Campus Plaine - CP 231, 1050 Bruxelles, Belgium
			\end{center}
		\end{minipage}
		}
	}
\newcommand{\uiuc}[1]{
	\centerline{
		\begin{minipage}[c]{0.7\textwidth}
			\begin{center}
			${}^{#1}$ Illinois Center for Advanced Studies of the Universe \& Department of Physics,\\ 
			University of Illinois, 1110 West Green St., Urbana IL 61801, U.S.A.
			\end{center}
		\end{minipage}
		}
	}

\begin{document}

\title{\thistitle}
\author{
	Luca Ciambelli$^{a}$ and Robert G. Leigh$^{b}$
	\\
	\\
	{\small \emph{\ulb{a}}} \\ \\
	{\small \emph{\uiuc{b}}}
	\\
	}
\date{}
\maketitle
\vspace{-0.5cm}
\begin{abstract}
\vspace{0.3cm}
A universal symmetry algebra organizing the gravitational phase space has been recently found. It corresponds to the subset of diffeomorphisms that become physical at corners -- codimension-$2$ surfaces supporting Noether charges. It applies to both finite distance and asymptotic corners. In this paper, we study this algebra and its representations, via the coadjoint orbit method. We show that generic orbits of the universal algebra split into sub-orbits spanned by finite distance and asymptotic corner symmetries, such that the full universal symmetry algebra gives rise to a unified treatment of corners in a manifold. We then identify the geometric structure that captures these algebraic properties on corners, which is the Atiyah Lie algebroid associated to a principal $GL(2,\RR)\ltimes \RR^2$-bundle. 
This structure is suggestive of the existence of a novel quantum gravitational theory which would unitarily glue such geometric structures, with spacetime geometries appearing as semi-classical configurations.
\end{abstract}

\begingroup
\hypersetup{linkcolor=black}
\tableofcontents
\endgroup
\noindent\rule{\textwidth}{0.6pt}

\setcounter{footnote}{0}
\renewcommand{\thefootnote}{\arabic{footnote}}
\newpage

\section{Introduction}

A complete understanding of symmetries in gauge theories, including gravity, is one of the key ingredients in the  description of the quantum properties of these theories. Since even in the same community the word 'symmetry' can have various meanings, let us clarify its meaning for us. A global symmetry of a theory expresses that there exists a conserved charge of motion, i.e., a quantity conserved dynamically. Classically, this is often seen at the level of the classical action for the theory, for instance a $U(1)$ global symmetry for a complex scalar field. From a quantum standpoint, one introduces a charge operator that acts on the Hilbert space of the theory, and conserved quantities are then interpreted as quantum numbers labeling states. 

Gauge symmetries on the other hand express mere redundancy of the description of a physical system. They express an underlying local reparameterization of the variables of the system, that leaves the latter invariant. The prime example for us is gravitational theories, where diffeomorphisms are local gauge symmetries. These gauge symmetries reduce the independent degrees of freedom of a given problem, that define the physical field space of the theory. 
As a result, the space of all fields is only a presymplectic manifold, by which one means that the gauge symmetries correspond to zero modes of a presymplectic form.
Global and gauge symmetries are at the core of Noether theorems \cite{Noether1918}: the first states that a global symmetry implies the existence of a conserved current, while the second states that conserved currents for gauge symmetries are weakly vanishing, that is, they vanish on  the equations of motion. Being conserved and weakly vanishing, a current associated to a gauge symmetry is at most a total derivative. Subtleties arise in the presence of boundaries, where this last statement opens the door to the possibility of having a non-trivial gauge symmetry current. This is the scenario we  explore in the context of diffeomorphism invariant theories.

In the presence of a boundary --- or, more generally, a subregion of interest --- part of the gauge symmetries acquires a non-trivial action on the field space approaching the boundary. Among bulk gauge symmetries, one must then distinguish between those that are still trivial from those that are no longer trivial. The former are still pure gauge symmetries while the latter become physical with non-zero charges.  They are not global symmetries of the full bulk field space, but nonetheless they are distinguished from the  gauge symmetries that remain pure gauge even in the presence of a boundary thanks to their non-vanishing Noether charge, which is now a codimension-$2$ integral from the bulk perspective, as first observed in the seminal work of Regge and Teitelboim \cite{Regge:1974zd}. These codimension-$2$ charges are called surface charges or, more recently, corner charges. The term corner refers to any codimension-$2$ surface on which these charges may have support. 
Corners, whether they be at finite or asymptotic distance, are then regarded as the building blocks of the \hlt{corner proposal} \cite{Donnelly:2016auv, Speranza:2017gxd,Geiller:2017whh,Freidel:2020xyx,Freidel:2020svx,Freidel:2020ayo,Donnelly:2020xgu,Ciambelli:2021vnn,Freidel:2021cjp,Ciambelli:2021nmv}, which is the starting point of this manuscript. We will return to a more thorough discussion of it presently. 

One of the main features of gauge theories is that they possess a  holographic feature: since an asymptotic symmetry gives rise to corner charges, one can reinterpret it as  a global symmetry of a theory supported on the codimension-$1$ boundary.  This result is perfectly in line with the AdS/CFT correspondence \cite{Maldacena:1997re,Witten:1998qj}, where the bulk asymptotic symmetries become the global conformal symmetries of the boundary field theory, as also precursorily observed in AdS$_3$ by Brown and Henneaux  \cite{Brown:1986nw}. It is by now more and more clear that gauge theories have in general this holographic nature, although we should emphasize that we do not imply that in any gauge theory the full bulk solution can be reconstructed from the boundary theory, but rather that the asymptotic field space and symmetries can always be reinterpreted in terms of a (not necessarily universal) boundary theory. Recent progress in this direction  notably includes flat holography \cite{Arcioni:2003xx,Arcioni:2003td,deBoer:2003vf,Dappiaggi:2005ci,Barnich:2010eb}, celestial holography \cite{Strominger:2013jfa,Kapec:2016jld,Cheung:2016iub,Pasterski:2016qvg,Pasterski:2017kqt,Donnay:2020guq,Pate:2019lpp,Fotopoulos:2019vac,Donnay:2021wrk,Pasterski:2021raf}, Carrollian holography \cite{Bagchi:2010zz,Hartong:2015xda,Ciambelli:2018wre, Ciambelli:2018ojf,Campoleoni:2018ltl, Figueroa-OFarrill:2021sxz,Donnay:2022aba}, but also finite distance boundaries, such as null hypersurfaces \cite{Hopfmuller:2016scf,Hopfmuller:2018fni,Chandrasekaran:2018aop,Ciambelli:2019lap,Speranza:2019hkr,Adami:2021nnf,Chandrasekaran:2021hxc} and in particular black hole horizons \cite{Donnay:2015abr,Donnay:2016ejv,Donnay:2019jiz,Grumiller:2019fmp,Carlip:2019dbu}. In \cite{Ciambelli:2021vnn}, we found that the subset of diffeomorphisms in gravitational theories that can contribute to asymptotic symmetries is universal. Due to the special role played by embeddings we called this algebra the maximal embedding algebra; we will take this opportunity to rebrand it as the \hlt{universal corner symmetry} (UCS), to emphasize its key role beyond classical embeddings. Indeed, we believe that this algebra should be taken as one of the fundamental ingredients of gravitational theories. 

The UCS is not realized fully in the vicinity of a single corner. Instead, we observe that only complementary subalgebras inside it are supported at different corners. The main distinction comes from finite- versus infinite-distance corners, from a bulk spacetime viewpoint. Finite distance corners realize a subalgebra called the \hlt{extended corner symmetry} (ECS) \cite{Donnelly:2016auv, Speranza:2017gxd,Freidel:2020xyx,Freidel:2020svx,Freidel:2020ayo,Donnelly:2020xgu,Ciambelli:2021vnn}, in which everything but Weyl, that is, arbitrary conformal rescalings of the corner, is in principle  sourced. This happens because the bulk metric approaches smoothly the corner, without poles in the normal directions, so the leading geometric structures are uncharged under Weyl. In contrast,  the algebra on asymptotic corners, i.e., corners at infinite metric distance, includes Weyl transformations, and a different subalgebra of the UCS is realized, which we name the \hlt{asymptotic corner symmetry} (ACS). The introduction of the ACS is a new result of this paper, and pertains both to asymptotically flat spacetimes, where the BMSW (Bondi-Metzner-Sachs-Weyl) or a subalgebra is realized\cite{Sachs:1962wk,doi:10.1098/rspa.1962.0161,doi:10.1098/rspa.1962.0206,Barnich:2010eb,Campiglia:2014yka,Compere:2018ylh,Campiglia:2020qvc,Flanagan:2019vbl,Freidel:2021fxf}, but also asymptotically AdS spacetimes, that have received tremendous attention in recent years with attempts to enlarge the asymptotic symmetry algebras \cite{Troessaert:2013fma,Grumiller:2016pqb,Ciambelli:2019bzz,Compere:2019bua,Alessio:2020ioh,Fiorucci:2020xto,Ciambelli:2020eba,Geiller:2022vto}. One of the main results of this paper is to clarify this universal structure and discuss how the ECS and ACS can be simultaneously realized inside the UCS. By focussing on this symmetry structure, this then allows for an interpretation of a corner purely group theoretically, without necessarily a reference to a classical spacetime in which it might be embedded. This is one of the first goals of the corner proposal and suggests a novel approach to quantum gravity.

With this in mind then, we focus attention on these symmetries of gravitational theories.
Especially in the field of asymptotic symmetries, most discussions follow a bulk-to-boundary (or bulk-to-corner) perspective, where the bulk and its classical dynamics are the starting point and the desired endpoint is a full characterization of the classical field space at a designated boundary. Given our interest in quantum aspects of gravity, we instead adopt a corner-to-bulk viewpoint, asking how this universal symmetry instructs us about gravity. We are not here looking for a holographic corner (or boundary) theory, but rather a full understanding of gravity itself from its `atomic' constituents, the corners. 

We have used classical gravity to identify the universal corner symmetry, but here we will start over with the UCS in hand but not necessarily an associated classical geometry. Our analysis will be presented in two steps. 
The first  is algebraic, and consists of understanding the representation theory of the UCS. To do so, we will apply the method of coadjoint orbits,\footnote{The coadjoint orbit method is becoming more and more utilized in our community \cite{Witten:1987ty,Balog:1997zz,Barnich:2014zoa,Barnich:2014kra,Duval:2014lpa,Barnich:2015uva,Donnelly:2020xgu,Barnich:2021dta,Lahlali:2021nrf,Marsot:2021tvq,Bergshoeff:2022eog,Riello:2022din} (see also \cite{Oblak:2016eij} and references therein).} developed by Kirillov in \cite{Kirillov_1962,
Kirillov1976ElementsOT,Kirillov_Merits,
Kirillov1990,kirillov2004lectures}, and further discussed in \cite{Kostant_2006,ginz,Duistermaat:1982aa,ALEKSEEV1989719,wildberger_1990,Brylinski_1994},  to the full universal corner symmetry.  
To this end, the Kirillov-Kostant-Souriau (KKS) symplectic two form \cite{souriau1970structure,Kostant_2006,
Kirillov1976ElementsOT,Kostant2009} is of central relevance, and we will find that the ECS and ACS emerge as important substructures of the orbits of the UCS. A key result is that the UCS does not possess Casimirs except those extracted from the corner reparameterization invariance, whereas the ACS and ECS do.  Locally, at a point on the corner, the analysis drastically simplifies, such that useful information can be extracted. The ECS and ACS at a point turns out to be ideals inside the UCS, and a full characterization of the tangent space to a generic point of the dual algebra is possible. 

The second step identifies the proper geometric framework to interpret the UCS orbit analysis. This involves the introduction of an Atiyah Lie algebroid, a concept that was originally described in \cite{Atiyah:1957,Atiyah:1979iu} (see also the extensive works of Mackenzie, \cite{mackenzie_1987,mackenzie_2005}), over the corner. Atiyah Lie algebroids have been advocated to be the mathematical foundation of gauge theories \cite{Lazzarini_2012,Fournel:2012uv,Jordan:2014uza,Carow-Watamura:2016lob,Kotov:2016lpx,Attard:2019pvw}. Here though we emphasize that the base manifold is not a spacetime, but instead a corner, such that the value of a local section of the algebroid is  given precisely by a generator of the full UCS. 
We recently explored in \cite{Ciambelli:2021ujl} the theory of algebroids, which finds a natural application in this context. We interpret the coadjoint analysis as a characterization of the dual Lie algebroid, where all the algebraic results derived locally can be imported in this more general setup.
The algebroid comes equipped with the adjoint representation, while other representations are realized by introducing associated bundles to the algebroid. In particular, we identify a certain rank-$2$ affine associated bundle that we implicate in what we interpret in terms of a reconstruction of a classical bulk geometry. That is, this affine bundle can be endowed locally with fibre coordinates that may be identifed with the two normal directions in such a classical geometry, and we show that derivations on the affine bundle naturally encode the local bulk metric data that are involved in the non-zero Noether charges. 
 This is thus the bulk classical representation. While it seems compelling that such classical geometry can emerge, we believe that the construction will be even more powerful in more general quantum gravitational contexts.
This is the beginning of our elaboration of the corner proposal.

Indeed, once the correct geometric structure underlying corners and UCS is introduced, the next item to explore is the relationship between a classical field space and the various orbits in the dual algebroid. This is made explicit through the moment map, described in general in \cite{Kostant_2006,souriau1970structure}. 
The moment map enters naturally in our algebroid, and it is strictly related to the construction of the associated affine bundle. There are two moment maps relevant for us: the moment map for finite distance corners into the dual of the ECS, and for asymptotic corners into the dual of the ACS. 
We discuss how to realize these moment maps inside the dual of the full UCS, such that it captures both images of these moment maps in a unified way. The associated affine bundle is crucial, because it gives the representation of the UCS that is realized in classical gravity around corners.

Although this paper is not about  the covariant phase space, it is important that we have recently found \cite{Ciambelli:2021nmv,Freidel:2021dxw}
a formalism where all diffeomorphisms are on an equal footing, and associated Noether charges are always integrable. Indeed, as a consequence of the original works of Wald et al. \cite{Lee:1990nz,Wald:1993nt,Iyer:1994ys,Wald:1999wa} (see also \cite{Compere:2019qed,Harlow:2019yfa}), and of Barnich and Brandt \cite{Barnich:2001jy}, surface charges are not in general integrable. The new formalism introduces embedding fields as a refinement in gravity of the general theory of edge modes explored in \cite{Freidel:2015gpa,Donnelly:2016auv,Speranza:2017gxd,Geiller:2017whh,Freidel:2020xyx,Freidel:2020svx,Freidel:2020ayo,Donnelly:2020xgu,Freidel:2021cjp}. While an active area of research \cite{Speranza:2022lxr,Carrozza:2022xut,Kabel:2022efn,Goeller:2022rsx}, our extended phase space considerably facilitates the construction of the moment map for finite distance corners, allowing to map the full field space to the dual algebra, and not only the part without symplectic flux. 

The paper is organized as follows. We first review in Section \ref{sec2} the main results outlined in \cite{Ciambelli:2021vnn}, and start the process of disentangling them from the classical bulk spacetime. In Section \ref{sec3} we focus on the group-theoretical aspects of the UCS at single points on the corners. In this simpler setup, we  exploit in Subsection \ref{sec3.1} the coadjoint orbit method to invert the KKS 2-form on generic points and thus have a full understanding of the algebra structure. We derive a general formalism that we use to find the KKS 2-form intrinsically for the ECS \ref{sec3.2} and ACS \ref{sec3.3}, at a point on the corner. We conclude this section discussing how to embed the two intrinsic analysis inside the full algebra. An intrinsic analysis of the various orbits of UCS at a single point is offered in Appendix \ref{appA}. Section \ref{sec4} is then devoted to the geometric properties of corners. The reparametrization symmetry of the corner is introduced in Subsection \ref{sec4.1}, where we pass from the algebraic viewpoint to the theory of algebroids, which is described in Subsection \ref{sec4.2}. The study of associated bundles allows us to identify the bulk spacetime representation. Eventually, we discuss how the concept of moment map is naturally encompassed in this geometric structure in Subsection \ref{sec4.3}, and explicitly construct the moment map for finite distance and asymptotic corners in 
Subsection \ref{sec4.4}. We conclude with a recap and future perspectives of this work in Section \ref{sec5}.

\section{Universal Corner Symmetry}\label{sec2}

In this section, we review previous works on corner symmetries with the goal of setting up notation and nomenclature. One of the goals of this paper is to emphasize the importance \cite{Ciambelli:2021vnn} of the group $Diff(S)\ltimes GL(2,\RR)\ltimes \RR^2$. As such, we will refer to the algebra associated to this group as the universal corner symmetry (UCS). We will often have occasion to refer to the group $GL(2,\RR)\ltimes \RR^2$ and we will generally call this group $H$, and its Lie algebra $\mh$. We will also be interested in certain subalgebras (actually, ideals) of $\mh$, which we will introduce shortly. First let us recall the defining representation of the UCS. 

In \cite{Ciambelli:2021vnn}, given a classical spacetime manifold $M$, we identified the UCS as a maximal closed subalgebra of $\dM$ associated with a codimension-2 embedded (or immersed) subspace $\phi:S\to M$. This was obtained by introducing local coordinates on $M$ near the subspace $y^M=(u^a,x^i)$, with $a=0,1$ and $i=1,...,n$, with $d=n+2$. Introducing intrinsic local coordinates $\sigma^\alpha$ ($\alpha=1,...,n$) on $S$, the embedding map may be described by $\phi : (u^a(\sigma),x^i(\sigma))$. We refer to the special case $\phi_0 : (u^a(\sigma)=0,x^i(\sigma)=\delta^i_\alpha\sigma^\alpha)$ as the trivial embedding, in which the bulk coordinates are adapted to those of $S$. 

An arbitrary vector field on $M$ can be written as $\un\xi=\xi^i(u,x)\un\pa_i+\xi^a(u,x)\un\pa_a$ in these coordinates. In the case of the trivial embedding,\footnote{Here we refer to the trivial embedding for convenience. In \cite{Ciambelli:2021vnn}, we also emphasized the importance of embeddings that are infinitesimally close to the trivial embedding. In such a case, the expansion in powers of $u^a$ still makes sense. } we can expand its component functions near $u^a=0$,
\beqn\label{exp1}
\xi^b(u,x)&=&\sum_{n=0} \frac{1}{n!}u^{a_1}...u^{a_n} \xi_{(n)}^b{}_{a_1...a_n}(x)=\xi_{(0)}^b(x)+\xi_{(1)}{}^b{}_{a_1}(x)u^{a_1}+{1\over 2}\xi_{(2)}{}^b{}_{a_1a_2}(x)u^{a_1}u^{a_2}+\dots
\\
\xi^i(u,x)&=&\sum_{n=0} \frac{1}{n!}u^{a_1}...u^{a_n} \xi_{(n)}^i{}_{a_1...a_n}(x)=\xi_{(0)}^i(x)+\xi_{(1)}^i{}_{a_1}(x)u^{a_1}+{1\over 2}\xi_{(2)}^i{}_{a_1a_2}(x)u^{a_1}u^{a_2}+\dots\label{exp2}
\eeqn
The UCS is obtained by restricting to vector fields for which these expansions are truncated at first order for $\xi^b$ and zeroth order for $\xi^i$,
\beq\label{subalgebragens}
\un{\xi}=\xi_{(0)}^k(x)\un\pa_k +\Big(\xi_{(0)}^a(x)+u^b\xi_{(1)}{}^a{}_b(x)\Big)\un\pa_a,
\eeq
and we use a hat to denote $\un{\hat\xi}_{(0)}=\xi_{(0)}^i(x)\un\pa_i$. 
Indeed, these vector fields close under the Lie bracket. That is, they 
satisfy the closed algebra
\beqn\label{LieMsub}
\big[\un{\xi}_1,\un{\xi}_2\big]
&=&\nonumber
\big[\un{\hat\xi}_{(0)1},\un{\hat\xi}_{(0)2}\big]^j\un\pa_j
\\&&\nonumber
+\Big[
\un{\hat\xi}_{(0)1}(\xi_{(0)2}^b)
-\un{\hat\xi}_{(0)2}(\xi_{(0)1}^b)
-\xi_{(1)1}{}^b{}_a\xi_{(0)2}^a
+\xi_{(1)2}{}^b{}_a\xi_{(0)1}^a
\Big]\un\pa_b
\\&& 
+u^c
\Big[-\big[\xi_{(1)1},\xi_{(1)2}\big]^b{}_c
+\un{\hat\xi}_{(0)1}(\xi_{(1)2}{}^b{}_c)
-\un{\hat\xi}_{(0)2}(\xi_{(1)1}{}^b{}_c)
\Big]\un\pa_b.
\eeqn
The term closure refers here to the $u^a$ expansion. The aforementioned algebra is the maximal finitely generated truncation of $\dM$, in the sense that if the vector fields have certain powers of $u^a$ only, then the Lie bracket of two such fields has, at best, the same powers of $u^a$ appearing in its expansion.

As discussed in \cite{Ciambelli:2021vnn}, one then observes that the generators are valued in the Lie algebra of the group
\beq\label{maximo}
\underbrace{Diff(S)}_\text{$\xi^j_{(0)}$}\ltimes\Big( \underbrace{GL(2,\RR)}_\text{$\xi_{(1)}{}^a{}_b$}\ltimes \underbrace{\RR^{2}}_\text{$\xi_{(0)}^a$}\Big).
\eeq
The semi-direct product structure is identified by associating terms in the Lie bracket as  
\beqn\label{LieMsubexpl}
\big[\un{\xi}_1,\un{\xi}_2\big]
&=&
\underbrace{\big[\un{\hat\xi}_{(0)1},\un{\hat\xi}_{(0)2}\big]^j\un\pa_j}_\text{$\dS$}
\nonumber\\&&
+\Big[
\underbrace{\un{\hat\xi}_{(0)1}(\xi_{(0)2}^b)
-\un{\hat\xi}_{(0)2}(\xi_{(0)1}^b)}_\text{$\dS$ acts on $\RR^2$}
+\underbrace{\xi_{(1)2}{}^b{}_a\xi_{(0)1}^a
-\xi_{(1)1}{}^b{}_a\xi_{(0)2}^a}_\text{$\mathfrak{gl}(2,\RR)$ acts on $\RR^2$}
\Big]\un\pa_b
\nonumber\\&&
+u^c
\Big[-\underbrace{\big[\xi_{(1)1},\xi_{(1)2}\big]^b{}_c}_\text{$\mathfrak{gl}(2,\RR)$}
+\underbrace{\un{\hat\xi}_{(0)1}(\xi_{(1)2}{}^b{}_c)
-\un{\hat\xi}_{(0)2}(\xi_{(1)1}{}^b{}_c)}_\text{$\dS$ acts on $\mathfrak{gl}(2,\RR)$}
\Big]\un\pa_b.
\eeqn
The algebra $\dS\loplus \big( \mathfrak{gl}(2,\RR)) \loplus \RR^2\big)$ of the group $Diff(S)\ltimes \big(GL(2,\RR)\ltimes \RR^{2}\big)$ is the UCS (in \cite{Ciambelli:2021vnn}, this was referred to as the maximal embedding algebra). The importance of this algebra lies in its universality;  in particular, we have not referred to any dynamical fields (such as a metric) or to any particular spacetime geometry. It has been identified with very few assumptions. 

In \cite{Ciambelli:2021vnn} (see also \cite{Speranza:2017gxd,Freidel:2020xyx,Freidel:2020svx,Freidel:2020ayo,Donnelly:2020xgu,Freidel:2021cjp}), it was also found that corners support (a subalgebra of) the UCS. By this we mean that only a subset of all of the diffeomorphism Noether charge densities\footnote{Here, the reader should recall that in the case of gauge symmetries, a Noether current is an exact form determined by an $n$-form ($n=d-2$) on $M$, which given a codimension-$2$ embedding $\phi:S\to M$ can be pulled back to a top form on $S$ and integrated to obtain the corresponding charge.} can pull back to non-zero values on a corner. Thus corner symmetries are important from this symmetry perspective as well; in particular most of $\dM$ is always pure gauge, with a subset realized as the residual global symmetry associated with the presence of a given corner.  For example, the Noether charges of the classical Einstein-Hilbert theory at finite distance corners are non-zero only for $Diff(S)\ltimes H_s$, where $H_s=SL(2,\RR)\ltimes \RR^2$. The associated algebra is known as the extended corner symmetry (ECS); we now believe that this is the maximal symmetry supported by Noether charges of any diffeomorphism-invariant theory on generic 'finite distance' corners. Nevertheless, we will see in this paper that it is the UCS which should be regarded as the primary symmetry of diffeomorphism invariant theories. Indeed, it is the UCS that allows a unified treatment of finite distance corners and 'asymptotic' corners. The latter support another important subalgebra of the UCS, the Lie algebra of the group
\beq
Diff(S)\ltimes H_w,\qquad H_w := W\ltimes \RR^2,
\eeq
which we refer to as the asymptotic corner symmetry (ACS). The group $Diff(S)\ltimes H_w$ in fact contains BMSW, recently shown to be the general asymptotic algebra \cite{Freidel:2021fxf} in asymptotically-flat geometries. Here by $W$ we mean the group $\RR\subset GL(2,\RR)$ consisting of the trace element of $GL(2,\RR)$. 

Therefore the ECS and ACS are both contained in the UCS. The latter is not itself realized in the vicinity of a single corner, but it contains both finite distance and asymptotic corner algebras as subalgebras. At a given point on the corner $S$, we can reduce our attention to the group $H$. Then, the algebras of the two subgroups $H_w=H/SL(2,\RR)$ and $H_s=H/W$, $\mh_w$ and $\mh_s$ respectively, are ideals inside $\mh$. We believe that this is an important feature. As we will see in the next section, it is this property that allows at each point on the dual space $\mh^*$ to identify $\mh_w$ or $\mh_s$ directions uniquely. This is why we should regard the UCS as the fundamental algebra, whose representation theory will dictate both finite distance and asymptotic distance physics.

Before we continue to the coadjoint orbit analysis, we set up some notation. We introduce a basis for the Lie algebra $\mh$ as $\left(\un t^a{}_b,\un t_a\right)$, with $a,b=0,1$. Then any element $\un\mu\in\mh$ can be written
\beq
\un\mu=\theta^a{}_b\un t^b{}_a+ b^a\un t_a,
\eeq
and thus we can regard $\un\mu\leftrightarrow (\theta^a{}_b,b^a)$. The Lie bracket on $\mh$ is given by (see Section \ref{sec3} for further details)
\beq
[\un\mu,\un\nu]=
-[\theta,\theta']^b{}_c\un t^c{}_b 
-(\theta^a{}_bb'{}^b-\theta'{}^a{}_bb^b)\un t_a,
\eeq
where $\un\nu\leftrightarrow (\theta'{}^a{}_b,b'{}^a)$.
The defining representation $\rho:\mh\to T\RR^2$ is of the form
\beq\label{defrep}
\rho(\un t^a{}_b)=u^a\un\pa_b,\qquad 
\rho(\un t_b)=\un\pa_b.
\eeq
If we denote the coordinates by $u^a=(u,\rho)$, then the $\mathfrak{gl}(2,\RR)=\mw \oplus \mathfrak{sl}(2,\RR)$ generators can be written
\beq
\un W=\tfrac12 (u\un\pa_u+\rho\un\pa_\rho),\qquad \un L_3=\tfrac12 (u\un\pa_u-\rho\un\pa_\rho),\qquad \un L_+=u\un\pa_\rho,\qquad \un L_-=\rho\un\pa_u.
\eeq
So $\un W$ is the trace of $\rho (\un t^a{}_b)$ while the $\mathfrak{sl}(2,\RR)$ generators are the three traceless elements. The Lie brackets are
\beq
[\un W,\un L_3]=0=[\un W,\un L_\pm],\qquad 
[\un L_3,\un L_\pm]=\pm \un L_\pm,\quad
[\un L_+,\un L_-]= 2\un L_3.
\eeq
We denote the translations   by
\beq
\un T_-=\un\pa_u,\qquad \un T_+=\un\pa_\rho,
\eeq
and we then have
\beqn
\left[\un W,\un T_\pm\right]=-\tfrac12\un T_\pm,\qquad
\left[\un L_3,\un T_\pm\right]=\pm\tfrac12 \un T_\pm,\qquad 
\left[\un T_+,\un T_-\right]= \un 0,
\\
\left[\un L_+,\un T_+\right]=\un 0=\left[\un L_-,\un T_-\right],\quad
\left[\un L_+,\un T_-\right]=-T_+,\qquad \left[\un L_-,\un T_+\right]=-T_-,
\eeqn
which is the algebra $(\mw\oplus \mathfrak{sl}(2,\RR)) \loplus \RR^2$. The algebra $\big(\dS\loplus (\mw\oplus \mathfrak{sl}(2,\RR))\big) \loplus \RR^2$ is then obtained by taking vector fields with coefficients that are arbitrary functions of the corner coordinates. 

For finite distance corners in $M$, as found in \cite{Ciambelli:2021vnn,Freidel:2020xyx} and previously discussed in \cite{Speranza:2017gxd,Donnelly:2020xgu}, the most general algebra dynamically realized\footnote{These statements apply to any diffeomorphism-invariant theory, \cite{Speranza:2017gxd}.} is the aforementioned extended corner symmetry $(\dS\loplus \mathfrak{sl}(2,\RR)) \loplus \RR^2$, obtained from the differential representation by dropping $\un W$. On the other hand, for asymptotic corners, it has been found in \cite{Freidel:2021fxf} that the  most general algebra dynamically realized in asymptotically flat spacetimes is the Weyl BMS algebra (BMSW), given by the $(\dS\loplus \mw) \loplus \RR$ subalgebra of the ACS, obtained by excluding the $\mathfrak{sl}(2,\RR)$ and $\un T_+$ generators. This algebra has numerous physically relevant subalgebras; the generalized BMS algebra $\dS\loplus \RR$ found in \cite{Campiglia:2014yka,Compere:2018ylh}, the extended BMS algebra found in \cite{Barnich:2010eb}, where $\dS$ is reduced to the locally well-defined conformal Killing vectors on $S$, and finally the original BMS algebra found in \cite{doi:10.1098/rspa.1962.0161,doi:10.1098/rspa.1962.0206}, where $\dS$ is reduced to the globally well-defined conformal Killing vectors on $S$.

We emphasize that the ACS algebra and ECS algebra are not nested into one another. Rather, they are distinct quotients, as mentioned above, of the UCS in which complementary subalgebras of $\mathfrak{gl}(2,\RR)$ are retained. While it was so far unclear why only the charges of ECS (ACS) have support on finite distance (asymptotic) corners in diffeomorphism-invariant theories, we will find an explanation of this fact below by studying the coadjoint orbits of the UCS: the tangent space to the orbit passing through a generic point can be split into ECS and ACS parts universally by enforcing the constancy of Casimirs along the orbit. Which Casimir to enforce is dictated by which subalgebra one wants to restrict to, and it is this at the end the physical input on the system. 

Ultimately, we are interested then in the representation theory of the UCS and its subalgebras. For this reason, we will not make further use of the defining differential representation given above, but work more abstractly. In fact, we will abandon an interpretation of the algebras as realized by vector fields in a classical spacetime. Instead we will eventually be led to a geometrically well-defined interpretation in terms of certain bundles over the space $S$. Of central importance is an Atiyah Lie algebroid over $S$ with structure group $H$, and the representation theory of the UCS can be explored by considering bundles associated to the Lie algebroid. In particular, we will introduce a rank-2 affine bundle over $S$ whose total space may be thought of as a local picture of a corner in a spacetime manifold. There are specific moment maps that involve the introduction of certain geometric structures on the affine bundle and map these to the ECS or ACS orbits.

\section{Coadjoint Orbit Method}\label{sec3}

In this section, we study the coadjoint orbits of the UCS, and its various subcases previously introduced. While  mostly mathematical,  this analysis has important repercussions in understanding gravity and its constituents, that is, the codimension-2 corners. While $\dS$ is an important part of the UCS, it turns out to be instructive to first consider the coadjoint orbits of $\mh$. Thus we will split our analysis into two parts: we begin by focussing our attention on $\mh$ and its ideals only, and later add back $\dS$ consistently.  

\subsection{The Algebra $\mh$}\label{sec3.1}

Given the Lie group $H$, its Lie algebra $\mh$ is a vector space. One can consider the dual vector space $\mh^*$, defined as
\beq
m\in \mh^*,\quad \un\mu\in \mh,\quad m: \mh\to \RR
,\qquad m(\un\mu)\in \RR.
\eeq
While the Lie algebra bracket on $\mh$ defines the adjoint action of $\mh$ on itself,
\beq
ad_{\un\mu}\un\nu := [\un\mu,\un\nu],
\eeq
there is a corresponding coadjoint action of $\mh$ on $\mh^*$, with the defining property
\beq\label{coad}
(ad_{\un\mu}^*m)(\un\nu)=-m(ad_{\un\mu}\un \nu).
\eeq
From this we can define the map $ad^*_{\un\mu} : \mh^*\to \mh^*$. 
These are all the essential ingredients one needs to study coadjoint orbits. 

Coadjoint orbits are usually introduced at the group level
\beq
O_m=\{g\in H \ \vert \ Ad^*_{g}m\},
\eeq
with $Ad^*$ the group coadjoint action, but it is sufficient for our purposes to use the local algebra version
\beq
{\cal O}_m=\{\un\mu\in \mh \ \vert \ ad^*_{\un\mu}m\},
\eeq
which is more suitable for the forthcoming discussion that also includes $\dS$. Indeed one can interpret the $ad^*$ action as exploring the tangent space to an orbit at the given point $m\in\mh^*$. Not all algebra elements act on $m$ in a non-trivial way. Those that do not define the stabilizer subalgebra
\beq
{\cal S}_m=\{\un\mu \in \mh \ \vert \ ad^*_{\un\mu}m=m\},
\eeq
 with associated stabilizer (or isotropy) group $S_m$. An important result is the isomorphism 
\beq
O_m\simeq H/S_m.
\eeq
Another important result, and the key reason why we focus on coadjoint orbits, is that, in very simple instances, classifying them gives information about the representations of the algebra \cite{Kirillov1976ElementsOT}. There is no known direct correspondence in general (we indeed refer to it as the orbit {\it method}), and it has limitations, but with such a complicated object as the UCS, we believe it is one of the best paths towards an understanding of  the algebra.

Crucially, one can perform a unified treatment of orbits and stabilizers using the KKS symplectic $2$-form \cite{souriau1970structure,
Kostant_2006}:\footnote{In an abuse of notation, we call KKS 2-form both the contracted and non-contracted objects, $\Omega_m(\un\mu,\un\nu)$ and $\Omega_m(.,.)$. The meaning and distinction will always be clear from context.}
\beq\label{KKS2f}
\Omega_m: {\cal O}_m\otimes {\cal O}_m\to \RR, \qquad \Omega_m(\un\mu,\un\nu):=\Omega_m(ad^*_{\un\mu}m,ad^*_{\un\nu}m)= m([\un\mu,\un\nu]).
\eeq
The notation $\Omega_m(\un\mu,\un\nu)$ is merely a shortform that we use for convenience.
By definition, the KKS $2$-form is non-degenerate and thus invertible when restricted to the orbits. This is due to the property of $\mh^*$ being a Poisson manifold, partitioned by symplectic leaves, on which the KKS form becomes non-degenerate. The analysis of the various orbits and different points in $\mh^*$ reduces then to a full characterization of $\Omega$. 

For the specific algebra $\mh$, this is a tractable problem, as it can be thought of as a simple extension of the familiar analysis of the Poincar\'e group (in $1+1$ dimensions). We denote a basis of $\mh$ as $\left(\un t^a{}_b,\un t_a\right)$ with $a,b=0,1$. The Lie brackets on $\mh$ are of the form
\beqn\label{Alg}
\left[\un t^a{}_b,\un t^c{}_d\right]
=
\delta^c{}_b\un t^a{}_d -\delta^a{}_d\un t^c{}_b,
\qquad
\left[\un t^a{}_b,\un t_c\right]=-\pi^a{}_c{}^d{}_b\un t_d,
\qquad
\left[\un t_b,\un t_c\right]=\un 0.
\eeqn
The values of the coefficients $\pi^a{}_d{}^c{}_b$ that determine the structure constants depend on the precise choice of group, because given such a choice, the $\un t^a{}_b$ may satisfy some conditions.\footnote{The notation is set up to apply generally to $GL(k,\RR)\ltimes \RR^k$ where $a,b=0,...,k-1$ (which appears in the context of codimension-$k$ embeddings), but we restrict attention here to $k=2$. The notation also applies to subgroups of $GL(2,\RR)$, for which we modify $\pi^a{}_b{}^c{}_d$ accordingly. We will see examples of this in forthcoming subsections.} For the case studied in this subsection, $GL(2,\RR)$, there are no such conditions, and  we have 
$\pi^a{}_c{}^d{}_b=\delta^a{}_c\delta^d{}_b$.

We will denote general elements $\mh\ni\un\mu=\theta^a{}_b\un t^b{}_a+b^a\un t_a$, $\un\nu=\theta'{}^a{}_b\un t^b{}_a+b'{}^a\un t_a$, etc. Thus we have
\beqn\label{genalg}
ad_{\un\mu}\un\nu=[\un\mu,\un\nu]=
-[\theta,\theta']^b{}_c\un t^c{}_b 
-(\theta^a{}_bb'{}^c-\theta'{}^a{}_bb^c)\pi^b{}_c{}^d{}_a\un t_d.
\eeqn
The dual vector space $\mh^*$ can be endowed with the dual basis $\{t^a{}_b,t^a\}$ satisfying
\beq
t^a{}_b(\un t^c{}_d)=\pi^a{}_d{}^c{}_b
,\qquad 
t^a{}_b(\un t_c)=0,\qquad
t^a(\un t^b{}_c)=0,\qquad 
t^a(\un t_b)=\delta^a{}_b.
\eeq
We denote a general element of $\mh^*$ as $m=J^a{}_bt^b{}_a+P_at^a$, and so we have
\beq
m(\un\mu)
=
\theta^b{}_aJ^a{}_b+b^aP_a.
\eeq
One can regard $(J^a{}_b,P_a)$ as the coordinates of the point $m$ in this basis.

The coadjoint action \eqref{coad} of $\mh$ on $\mh^*$ satisfies
\beq
(ad^*_{\un\mu}m)(\un\nu)=-m(ad_{\un\mu}\un\nu)
=J^a{}_b [\theta,\theta']^b{}_a
+ P_a(\theta^a{}_bb'{}^b-\theta'{}^a{}_bb^b),
\eeq
from which it is then a simple matter to deduce
\beq\label{adactionhdual}
ad^*_{\un\mu}m=
\Big([J,\theta]^a{}_b - b^aP_b\Big) t^b{}_a
+P_b\theta^b{}_ct^c.
\eeq
Given the explicit expression of $m$,
one can interpret the $ad^*$ action as a transformation of its components,
\beqn\label{dJdP}
\delta_{\un\mu} J^a{}_b
= [J,\theta]^a{}_b  -b^aP_b,
\qquad
\delta_{\un\mu} P_a = P_b\theta^b{}_a.
\eeqn
Here $\delta_{\un\mu} J^a{}_b$ and $\delta_{\un\mu} P_a$ should be regarded as components of tangent vectors to the orbit at the point $m$.

A relevant intermediate step in the calculation of the symplectic form is that we can invert these relations to obtain
\beqn\label{inversions}
b^a&=&
-\delta_{\un\mu} J^a{}_b J^b{}_c \kappa^{cd}P_d
-J^a{}_b \delta_{\un\mu} J^b{}_c \kappa^{cd}P_d
+{\cal J} \delta_{\un\mu} J^a{}_b\kappa^{bc}P_c, 
\\
\theta^a{}_b&=&-\delta_{\un\mu} J^a{}_c \kappa^{cd} P_d P_b+\big(P_e\delta_{\un\mu} J^e{}_c \kappa^{cd} P_d\big) \delta^a{}_b+\delta_{\un\mu} P_c \kappa^{cd}P_d J^a{}_b
-P_eJ^e{}_d\kappa^{cd}\delta_{\un\mu} P_c \delta^a{}_b.
\label{inversionsb}
\eeqn
In this expression, we have introduced the trace ${\cal J}=J^a{}_b\delta^b{}_a$, and we have defined the $\mh$-invariant quantity
\beq\label{C3}
\kappa^{ab}={\varepsilon^{ab}\over C_3}, \qquad C_3:= P_aJ^a{}_b\varepsilon^{bc}P_c.
\eeq
The symbol $\varepsilon$ is the $2x2$ Levi-Civita symbol with conventions $\varepsilon^{01}=-1$, and we note that $\varepsilon^{ab}$ and $C_3$ are $\mh_s$ invariants.
We regard $C_3$ as an element of the enveloping algebra of $\mh^*$. The facts that $b^a$ depends only on $\delta_{\un\mu}J^a{}_b$ while $\theta^a{}_b$ depends on both $\delta_{\un\mu}J^a{}_b$ and $\delta_{\un\mu}P_a$ are a result of the semi-direct sum structure of the algebra. The algebra $\mh$ has dimension $6$. That the inversion \eqref{inversions} and \eqref{inversionsb} are possible is testament to the fact that the orbit at a generic point $m$ can be $6$-dimensional. On the other hand, 
the inversion formulas fail if $C_3$ vanishes, indicating that the special points that satisfy this condition have lower dimensional orbits.  This is the sense in which $\mh^*$ fails to be a symplectic manifold, as there are a series of non-generic points for which the orbits are 4-,2- or 0-dimensional. We refer to such orbits as singular orbits, and explore them in Appendix \ref{appA}. Clearly $C_3$ plays a central role here, and we will return to it presently.

As a final step, we write the KKS symplectic form
\beqn\label{KKS}
\Omega^{(\mh)}_m(\un\mu,\un\nu)
=
-J^a{}_b[\theta,\theta']^b{}_a  -P_a(\theta^a{}_cb'{}^c-\theta'{}^a{}_cb^c) ,
\eeqn
which is a non-degenerate 2-form on the orbit. 
Assuming $C_3\neq 0$, we can now use the inversion formulas \eqref{inversions} and \eqref{inversionsb} to rewrite this in terms of $\delta_{\un\mu} J^a{}_b$ and $\delta_{\un\mu} P_a$. It is convenient to first separate the trace and traceless pieces,
\beq
J^a{}_b=\bar J^a{}_b+\tfrac12 {\cal J}\delta^a{}_b,\qquad 
\theta^a{}_b=\bar\theta^a{}_b+\tfrac12 w\delta^a{}_b,\qquad w=\theta^a{}_b\delta^b{}_a,
\eeq
and we then obtain 
\beqn
\Omega^{(\mh)}_m(\un\mu,\un\nu)=\delta_{\un\mu}P_a\big(\delta_{\un\nu}\bar J^a{}_b\bar J^b{}_c+\bar J^a{}_b \delta_{\un\nu}\bar J^b{}_c\big)\kappa^{cd}P_d
-\tfrac{1}{2}(P_a\delta_{\un\mu}\bar J^a{}_b)\kappa^{bc}(P_d\delta_{\un\nu}\bar J^d{}_c)
+\tfrac12\delta_{\un\mu}\log C_3\delta_{\un\nu}{\cal J}
-(\un\mu \leftrightarrow \un\nu).\label{KKSdJdP}
\eeqn
To derive this, we have also used the  identity 
\beq 
\delta_{\un\mu} \log C_3=w,
\eeq 
which follows from \eqref{inversionsb}.
We note that the Pfaffian of $\Omega^{(\mh)}_m$ is 
\beq \label{pfaf}
Pf[\Omega^{(\mh)}_m]=-\frac{1}{C_3},
\eeq 
which is further evidence that the KKS form is non-degenerate iff $C_3\neq 0$. In the formula \eqref{KKSdJdP}, there are terms involving only the traceless part of $J^a{}_b$ (and thus associated with $\mh_s$). The last  term instead indicates that the trace ${\cal J}$ is canonically conjugate to $\log C_3$.  

Now if we denote $Z^A=(J^a{}_b,P_a)$ and write the symplectic 2-form as \[\Omega_m(\un\mu,\un\nu)= \Omega_{AB}(Z)(\delta_{\un\mu}Z^A\delta_{\un\nu}Z^B-\delta_{\un\nu}Z^A\delta_{\un\mu}Z^B),\] then by inverting the matrix $\Omega_{AB}(Z)$, we obtain Poisson brackets as usual $\{Z^A,Z^B\}=\Omega^{AB}(Z)$. In this case, we find simply the Poisson bracket representation of the full $\mh$ algebra, as long as $C_3\neq 0$. 

Let us recap our findings so far. The central result is that \eqref{KKS} at the generic point $m$ can be inverted to then obtain the KKS symplectic form on $6$-dimensional orbits, giving the Poisson bracket representation of $\mh$. For such orbits, there is no isotropy group and no Casimirs. This is achievable only if $C_3\neq 0$, indicating that $\mh^*$ is locally symplectic, but there may be lower dimensional singular orbits if $C_3=0$, as we show in Appendix \ref{appA}.   There is a completely different structure arising in the case $\mh_s^*$ and $\mh_w^*$, where we will find that generic orbits have always non-trivial isotropy groups. As already discussed previously, one can single out $\mh_s$ and $\mh_w$ directions at generic $\mh^*$ points. We will show this after presenting an intrinsic analysis of $\mh_s$ and $\mh_w$ coadjoint orbits. 

\subsection{The Algebra $\mh_s$}\label{sec3.2}

We begin with the intrinsic derivation of the orbits of $\mh_s$. Our general analysis is ready-made for such a scenario. Indeed, it suffices to consider \eqref{Alg},
\beqn
\left[\un t^a{}_b,\un t^c{}_d\right]
=
\delta^c{}_b\un t^a{}_d -\delta^a{}_d\un t^c{}_b,
\qquad
\left[\un t^a{}_b,\un t_c\right]=-\pi^a{}_c{}^d{}_b\un t_d,
\qquad
\left[\un t_b,\un t_c\right]=\un 0,
\eeqn
with the choice $\pi^a{}_c{}^d{}_b=\delta^a{}_c\delta^d{}_b-\tfrac12 \delta^a{}_b\delta^d{}_c$. Then one obtains
\beq
m(\un\mu)
=
\theta^b{}_aJ^a{}_b-\tfrac12 {\cal J}w+b^aP_a=\bar\theta^b{}_aJ^a{}_b+b^aP_a,
\eeq
where we recall that $\bar\theta^a{}_b=\theta^a{}_b-\tfrac12 w\delta^a{}_b$ is the traceless part of $\theta^a{}_b$. We see that the trace $w$ decouples, indicating that we are appropriately describing the algebra $\mh_s$. We also see that consistently only the traceless parts of $J^a{}_b$ appear here (since only those coordinatize $\mh_s^*$), and so we will write it as $\bar J^a{}_b$. The KKS form then reads
\beqn
\Omega^{(\mh_s)}_m(\un\mu,\un\nu)
&=&
-J^a{}_b[\theta,\theta']^b{}_a  -P_a(\theta^a{}_cb'{}^c-\theta'{}^a{}_cb^c)+\tfrac12 P_a(w b'{}^a-w' b^a)\nonumber\\
&=&
-\bar J^a{}_b[\bar\theta,\bar\theta']^b{}_a  -P_a(\bar\theta^a{}_cb'{}^c-\bar\theta'{}^a{}_cb^c).\label{intKKS}
\eeqn
Again, we see that only the traceless quantities contribute to the $\mh_s$ symplectic form.
Working through a similar analysis as in the last subsection leads to the following expressions for the variations corresponding to the $ad^*$ action of $\mh_s$ on $\mh_s^*$,
\beqn\label{var2}
\delta_{\un\mu}\bar J^a{}_b
= [\bar J,\bar \theta]^a{}_b  -b^aP_b+\tfrac12 b^cP_c\delta^a{}_b,
\qquad
\delta_{\un\mu} P_a = P_b\bar\theta^b{}_a.
\eeqn
One observes that $\delta^b{}_a \delta_{\un\mu}J^a{}_b=0$, and therefore there are only $3$ variations inside $\delta_{\un\mu}J^a{}_b$, which are the traceless parts, $\delta_{\un\mu}\bar J^a{}_b$.
While the algebra has dimension $5$, the tangent space at a generic point in $\mh_s^*$ is only $4$-dimensional (consistent with such orbits being symplectic). This can be seen by attempting to invert \eqref{var2}. While there are five equations depending on the five parameters $(\bar\theta^a{}_b,b^a)$, they are not all independent. Indeed one finds that the variations satisfy a relation independent of  $(\bar\theta^a{}_b,b^a)$,
\beq\label{P0v}
2(\bar J^0{}_1P_0-\bar J_3P_1)\delta_{\un\mu}P_0=2(P_1\bar J^1{}_0+\bar J_3 P_0)\delta_{\un\mu}P_1-P_0^2\delta_{\un\mu}\bar J^0{}_1+2P_0P_1\delta_{\un\mu}\bar J_3+P_1^2\delta_{\un\mu}\bar J^1{}_0,
\eeq
where for brevity we have introduced 
$\bar J_3=\bar J^0{}_0=-\bar J^1{}_1$.
This result is easy to understand: since we are considering here $\mh_s$, we have that $C_3$ is a Casimir. Indeed, regarding $\delta_{\un\mu}$ as a derivation, we readily compute
\beqn
\delta_{\un\mu}C_3&=&\delta_{\un\mu}(P_a\bar J^a{}_b\varepsilon^{bc}P_c)\nonumber\\
&=&2\delta_{\un\mu}P_a\,\bar J^a{}_b\varepsilon^{bc}P_c
+P_a\delta_{\un\mu}\bar J^a{}_b\varepsilon^{bc}P_c\nonumber
\\
&=&2\delta_{\un\mu}P_0(\bar J^0{}_1P_0-\bar J_3P_1)
-2\delta_{\un\mu}P_1(\bar J_3 P_0+\bar J^1{}_0P_1)
+P_0^2\delta_{\un\mu}\bar J^0{}_1
-2P_0P_1\delta_{\un\mu}\bar J_3
-P_1^2\delta_{\un\mu}\bar J^1{}_0.
\eeqn
The vanishing of the variation of $C_3$ is thus equivalent to the relation \eqref{P0v}: since $C_3$ is a Casimir, it is constant along an $\mh_s$ orbit. 

Given that, we can obtain the KKS form on such an orbit by using the above relation to eliminate one of the variations. This cannot be done globally, and there is in general some freedom of choice. As an example, suppose we are at a point $m\in\mh_s^*$ where $\bar J^0{}_1P_0-\bar J_3P_1\neq 0$. We could then choose to eliminate $\delta_{\un\mu}P_0$ by solving \eqref{P0v} for it as
\beq\label{P0v2}
\delta_{\un\mu}P_0=\frac{2(P_1\bar J^1{}_0+\bar J_3 P_0)\delta_{\un\mu}P_1-P_0^2\delta_{\un\mu}\bar J^0{}_1+2P_0P_1\delta_{\un\mu}\bar J_3+P_1^2\delta_{\un\mu}\bar J^1{}_0}{2(\bar J^0{}_1P_0-\bar J_3P_1)}.
\eeq
If we then define
\beq
J_C^2:=\bar J^a{}_b\bar J^b{}_a=2(\bar J_3^2+\bar J^1{}_0\bar J^0{}_1), 
\eeq
we obtain from \eqref{intKKS} the non-degenerate KKS form on the $4$-dimensional orbits
\beqn\label{KKSintS}
\Omega^{(\mh_s)}_m(\un\mu,\un\nu)
=\frac{2P_0\delta_{\un\mu}\bar J_3\delta_{\un\nu}\bar J^0{}_1+P_1\delta_{\un\mu}\bar J^1{}_0\delta_{\un\nu}\bar J^0{}_1+\delta_{\un\mu}P_1\delta_{\un\nu}J_C^2}{2 (\bar J^0{}_1P_0-\bar J_3P_1)}-(\un\mu \leftrightarrow \un\nu).
\eeqn
The quantity $J_C^2$ is of course an $\mathfrak{sl}(2,\RR)$ Casimir; it is not on the other hand an $\mh_s$ Casimir and consequently varies along the orbit. Inverting the components of this symplectic form gives (basis ordering $(P_1,\bar J_3,\bar J^0{}_1,\bar J^1{}_0)$)
\beq
\Omega_{AB}^{(\mh_s)}=\begin{pmatrix}0&\tfrac12 P_1&0&-P_0\cr-\tfrac12 P_1&0&-\bar J^0{}_1&\bar J^1{}_0\cr 0&\bar J^0{}_1&0&-2\bar J_3\cr P_0&-\bar J^1{}_0&2\bar J_3&0\end{pmatrix},
\eeq
and thus gives rise to the Poisson bracket realization of $\mathfrak{sl}(2,\RR)$ along with the appropriate brackets of $P_1$ with $\bar J^a{}_b$. We reiterate that although we are not here using $\delta P_0$ as a basis form, $P_0$ does vary along the orbit.  Thus the full $\mh_s$ algebra is encoded in the Poisson  brackets if we recall that $\delta C_3$ is normal to the orbits.

To recap, the intrinsic $\mh_s$ analysis has revealed that generic orbits are $4$-dimensional, and thus there is a constraint among the $5$ variations. We have seen that this is a result of the existence of an $\mh_s$ cubic Casimir, $C_3$,  which can then be used to express one variation in terms of the others. Solving this for $\delta_{\un\mu}P_0$, one obtains exactly \eqref{P0v}. Enforcing the Casimir $C_3$ to be constant is indeed what we will do in order to see $\mh_s$ directions directly inside the $6$-dimensional tangent space of $\mh$ orbits in Subsection \ref{sec3.4}. First, we will in the next subsection discuss the intrinsic analysis for  $\mh_w$ orbits. 

\subsection{The Algebra $\mh_w$}\label{sec3.3}

A similar intrinsic analysis can be performed to find generic $\mh_w$ orbits. Here again our setup is ready to be applied. We use \eqref{Alg}, i.e.,
\beqn
\left[\un t^a{}_b,\un t^c{}_d\right]
=
\delta^c{}_b\un t^a{}_d -\delta^a{}_d\un t^c{}_b,
\qquad
\left[\un t^a{}_b,\un t_c\right]=-\pi^a{}_c{}^d{}_b\un t_d,
\qquad
\left[\un t_b,\un t_c\right]=\un 0,
\eeqn
with now $\pi^a{}_c{}^d{}_b=\tfrac12 \delta^a{}_b\delta^d{}_c$. This choice results in only the trace of the generators $\un t^a{}_b$ appearing. It immediately follows that
\beq
m(\un\mu)
=\tfrac12 {\cal J}w+b^aP_a,
\eeq
and the KKS form then reads
\beqn
\Omega^{(\mh_w)}_m(\un\mu,\un\nu)
=-\tfrac12 P_a(w b'{}^a-w' b^a).
\eeqn
Again, we see that  the traceless $\bar\theta^a{}_b$ does not contribute to the intrinsic $\mh_w$ analysis, with only the trace $w$ appearing. 
Similar to previous discussions, we read off the variations
\beqn\label{var3hw}
\delta_{\un\mu} {\cal J}
= -b^cP_c,
\qquad
\delta_{\un\mu} P_a = \tfrac12 w P_a,
\eeqn
corresponding to the $ad^*$ action of $\mh_w$ on $\mh_w^*$.
Here the algebra has dimension $3$, but the tangent space to an orbit at a generic point in $\mh_w^*$ will only be $2$-dimensional. Indeed, we see from the second two equations in \eqref{var3hw} that $w$ may be eliminated giving an equation amongst the variations. At all points in $\mh_w$ for which $P_0\neq 0$, we can write this equation as
\beq\label{solveCas1}
\delta_{\un\mu}P_1=P_1 {\delta_{\un\mu}P_0\over P_0}.
\eeq
Using this to eliminate $\delta_{\un\mu}P_1$ allows us to extract the non-degenerate KKS form on the $2$-dimensional orbits,
\beqn\label{KKSintW}
\Omega^{(\mh_w)}_m(\un\mu,\un\nu)={\delta_{\un\mu}P_0\delta_{\un\nu}{\cal J}-\delta_{\un\mu}{\cal J}\delta_{\un\nu}P_0\over  P_0},
\eeqn
which is valid on the domain $P_0\neq 0$. We see that ${\cal J}$ and $\log P_0$ are Darboux coordinates on these 2-dimensional orbits. Note that eq. \eqref{solveCas1} can be interpreted as the vanishing of an $\mh_w$ Casimir, which is of the form
\beq
C_1={P_1\over P_0}.
\eeq
Indeed, since the $P_a$ simply rescale with the same weight under the Weyl transformation, their ratio is invariant, and thus each orbit has a constant value of $C_1$. 
In the following subsection, we will see how to realize $\mh_s$ and $\mh_w$ directions directly inside the $6$-dimensional tangent space on $\mh^*$ by making use of the facts gleaned from the intrinsic analyses.

\subsection{Orbits and Ideals of $\mh$}\label{sec3.4}

The algebra $\mh$ contains three ideals: an Abelian one, $\mh_0\equiv \RR^2$, and the two non-Abelian ideals, $\mh_s$ and $\mh_w$, whose orbits we considered above. It is not a coincidence that the two non-Abelian ideals are exactly the algebra of finite distance and asymptotic corners. Indeed, the property of being ideals of $\mh$ implies that one can reach them from  a quotient of the original group $H$:
\beq
H_0=H/GL(2,\RR), \qquad H_w=H/SL(2,\RR),\qquad H_s=H/W.
\eeq

In this section, we will discuss how the $6$-dimensional tangent space to a generic point in $\mh^*$ can be regarded as containing complementary subspaces that can be associated to the orbits of $\mh_w^*$ and $\mh_s^*$, respectively. 

We have of course an action of $\mh$ on $\mh^*$. We can ask how we might reduce to an action of the ideals $\mh_s$ and $\mh_w$ in a geometrically meaningful way. We can make progress in this direction by noting that the two ideals have Casimirs which are local functions on $\mh^*$. In the case of $\mh_s$, there is one such Casimir $C_3$ that we have seen in the above discussions, while for $\mh_w$ there are three Casimirs that we will introduce momentarily. 

Let us first discuss the case of $\mh_s\subset\mh$. Formally, we can introduce the distribution $ker(\delta C_3)$, which we regard as a locally integrable distribution in $T\mh^*$; identifying $T_m\mh^*$ with $\mh$, this consists of all vectors $\un\mu$ that satisfy $\delta_{\un\mu}C_3=0$. Since 
we know that $\delta_{\un\mu}C_3=wC_3$, we see that $\delta_{\un\mu}C_3$ vanishes when $w=0$, and so we can identify $ker(\delta C_3)\sim \mh_s$. 
This defines a $5$-dimensional subspace of $T_m\mh^*$.  We know from eq. \eqref{KKSdJdP} that on this subspace, the KKS form is singular and reduces to that of the intrinsic $\mh_s$ analysis, eq. \eqref{KKSintS}. This indicates that a $4$-dimensional subspace of $ker(\delta C_3)$ can be associated with an $\mh_s$ orbit. However, if we examine the action of $\mh_s$ on $\mh^*$, we find from \eqref{dJdP}
\beqn
\delta_{\un\mu}\bar J^a{}_b
= [\bar J,\bar\theta]^a{}_b  -b^aP_b+\tfrac12 b^c P_c \delta^a{}_b ,
\qquad
\delta_{\un\mu} P_a = P_b\bar\theta^b{}_a,
\qquad
\delta_{\un\mu}{\cal J}
=   -b^c P_c.
\eeqn
In this expression, we recover the intrinsic result \eqref{var2}. Moreover, whereas the trace ${\cal J}$ decouples from the KKS form when pulled back to $ker(\delta C_3)$, it is not invariant under $\mh_s$. This indicates that the orbit is not just a constant-${\cal J}$ slice through $\mh^*$. 

Similarly, one can identify $\mh_w\subset \mh$ by constructing the $3$-dimensional distribution $ker(\delta C_1,\delta C_2,\delta C_2')$, where 
\beq
C_1:={P_1\over P_0},\qquad C_2:= C_1^{-1}J^0{}_1-J^0{}_0,\qquad C_2':= C_1J^1{}_0-J^1{}_1.
\eeq
These three functions have been chosen such that $ker(\delta C_1,\delta C_2,\delta C_2')$ consists of those vectors $\un\mu$ for which $\bar\theta^a{}_b=0$. Indeed, calling $\theta_3={1\over 2}(\theta^0{}_0-\theta^1{}_1)$, one finds
\beqn
\delta_{\un\mu}\begin{pmatrix}C_1\cr C_2\cr C_2'\end{pmatrix}
&=& \begin{pmatrix}1&-C_1^2&-2C_1\cr
(C_2'-C_2)/C_1&0&0\cr 0&-(C_2'-C_2)C_1&0\end{pmatrix}
\begin{pmatrix}\theta^0{}_1\cr\theta^1{}_0\cr \theta_3\end{pmatrix}.
 \eeqn
Since the determinant of the matrix appearing here is given by $2(C_2'-C_2)^2C_1$, generically the left-hand side vanishes only for $\bar\theta^a{}_b=0$, as required. Thus we conclude that $ker(\delta C_1,\delta C_2,\delta C_2')\sim \mh_w$, and $C_1,C_2,C_2'$ are Casimirs of $\mh_w$. In terms of these variables, the KKS form on $\mh^*$ can be written 
\beqn
\Omega^{(\mh)}_m(\un\mu,
\un\nu)=
-\delta_{\mu}\log C_1 \delta_{\nu} J_3
+\delta_{\mu} C_2\delta_{\nu}\log (C_2'-C_2)
+\tfrac12\delta_{\mu}\log (C_1P_0^2)\delta_{\nu}{\cal J}-(\un\mu\leftrightarrow\un\nu).
\eeqn
Thus we immediately see that this pulls back to $\delta\log P_0\wedge \delta {\cal J}$, which coincides with the KKS form on $\mh_w^*$. Similarly to what happens for $\mh_s$, while $J_3$ decouples from the KKS form, it is not invariant under $\mh_w$. Indeed, from \eqref{dJdP} reduced to $\mh_w$, we have
\beqn
\delta_{\un\mu} J_3
=   -\tfrac12(b^0-b^1C_1)P_0,
\qquad
\delta_{\un\mu} {\cal J}
=   -(b^0+b^1C_1)P_0,
\qquad
\delta_{\un\mu}\log P_0 = \tfrac12 w.
\eeqn

Finally, we note that the reduction to $\mh_0$ can be performed by considering the distribution corresponding to setting $\bar\theta^a{}_b=0$ and $w=0$; this however yields 
\beq
\Omega^{(\mh)}_m(\un\mu,\un\nu)=0,
\eeq
and the isotropy algebra coincides with the full algebra itself.

So we have seen that the three ideals of $\mh$ give rise to $4$-, $2$-, and $0$-dimensional immersions of their orbits at generic points. On the other hand, the generic orbits of $\mh$ itself are $6$-dimensional, and indeed there are no Casimirs. Casimir operators, and related quantum numbers, are at the core of the quantum representation theory, and we plan to address quantum features of this analysis in future publications. 

We can now continue the discussion of the beginning of this subsection. We have shown that the $6$ directions $\{\delta_{\un\mu}P_a,\delta_{\un\mu}J^a{}_b\}$ are all independent at a generic point in $\mh^*$. Furthermore, we have seen that the orbits of the ideals in $\mh$ can be regarded as immersed in $\mh^*$. 
The algebra $\mh$, and more generally the full UCS, serves as an organizing principle for the orbits of $\mh_s$ and $\mh_w$, which we have argued correspond to the physically relevant finite distance and asymptotic corners, respectively. More precisely, we should lift this analysis to the ECS and ACS, by including as well the semi-direct product with $\dS$.
A full accounting of the coadjoint orbits of UCS, ECS and ACS is beyond the scope of this paper. However, we believe that the simplified discussion given above of $\mh,\mh_s$ and $\mh_w$ is helpful for organization. In the next section, we will add back in $\dS$ and thus discuss the full UCS. We will find that this can be usefully organized in terms of certain Lie algebroids over $S$. 

Importantly, our analysis will suggest that there may be a useful semi-classical construction of spacetime geometries making use of these algebroid structures that have support on corners. To make such a contact with classical physics, we will build moment maps pertaining to orbits of the ECS and ACS. 

\section{UCS and Its Algebroid Interpretation}\label{sec4}

Restoring the $\dS$ part of the UCS has dramatic effects on the coadjoint analysis. After discussing them, we will show how algebroids offer the natural playground to geometrically describe the UCS over a corner. Associated bundles to the algebroid give various representations of the UCS, including what we identify as the classical spacetime representation. We then reach the right point to introduce moment maps, and discuss how the dual of the UCS algebroid contains the image of both the ECS and ACS moment maps.

\subsection{Reintroduction of  $\dS$}\label{sec4.1}

We would like now to reintroduce the $\dS$ part of the algebra. Given that it acts on everything to its right, but it is not acted upon, due to the semi-direct structure, the analysis carried so far goes through, with however the important modification that elements on $\mh$ are now valued on $S$. In particular, the UCS, ECS, and ACS are obtained from $\mh$, $\mh_s$, and $\mh_w$ by adding $\dS$ and making the latter algebras local, respectively. The UCS is the biggest, and the ECS and ACS are subalgebras inside it. The ECS and ACS are not however ideals, because $\dS$ acts non-trivially on the complementary elements of the UCS in these two reductions. This has important repercussions on the KKS form, that we detail below.

In this context, the dual space UCS$^*$ is coordinatized by a one-form $\alpha=\alpha_\beta(\sigma)d\sigma^\beta$ on $S$ together with an element of $\mh^*$ with values in $S$, that is, calling $M\in$ UCS$^*$,
\beq
M=\alpha_\beta(\sigma) d\sigma^\beta+J^a{}_b(\sigma) t^b{}_a+ P_a(\sigma) t^a.
\eeq
Similarly, an element $\un\mX\in $UCS is given by a vector field on $S$ together with an element of $\mh$ valued on $S$,
\beq
\un\mX=\xi^\beta(\sigma)\un\pa_\beta+\theta^a{}_b(\sigma)\un t^b{}_a+ b^a(\sigma)\un t_a.
\eeq
One can then act with $M$ on $\un\mX$ producing a function on $S$ which evaluates to
\beq
M(\un\mX)=i_{\un\xi}\alpha+\theta^a{}_bJ^b{}_a+b^aP_a,
\eeq
where the explicit $\sigma$ dependence is from now on implicitly assumed, and we have introduced the interior product $i_{\un\xi}$ on $S$.
The invariant pairing $\langle \ , \rangle:$ UCS$^*\otimes$ UCS $\to \RR$ is then introduced as
\beq\label{ucspair}
\langle M,\un\mX\rangle=\int_S vol_S\, M(\un\mX),
\eeq
where $vol_S$ is a volume form on the corner $S$. Once a field space is introduced, $M(\un\mX)$ will be the image in UCS$^*$ of a local charge aspect via a  moment map. On the other hand, the invariant pairing is the image under the moment map of an integrated charge on the corner. We will discuss this further in Subsection \ref{sec4.3}.

Now, the adjoint action of the Lie algebra UCS on itself is given by
\beq
ad_{\un\mX}\un\mY:= [\un\mX,\un\mY].
\eeq
Correspondingly, the coadjoint action  is defined through the pairing given above,
\beq\label{adad}
\langle ad^*_{\un\mX}M,\un\mY\rangle =- \langle M, ad_{\un\mX}\un\mY\rangle.
\eeq
Thanks to this pairing, we extend the coadjoint computations on $\mh^*$ to UCS$^*$, with some important limitations. Indeed, as we will discuss further in the subsequent section, we will interpret a generator of the UCS geometrically as a section of an algebroid rather than an algebra. While in the latter we can descend to a group analysis, one should generalize certain results to groupoids in order to do so here. We expect this to be a promising avenue of research, but would go beyond the scope of this paper.

Given $M$ in UCS$^*$, we can define from \eqref{adad} the map $ad^*_{\un\mX} :$ UCS$^*\to$ UCS$^*$. Elements of the UCS whose coadjoint action is trivial defines the stabilizer subalgebroid
\beq
{\cal S}_M=\{\un\mX \in \text{UCS} \ \vert \ ad^*_{\un\mX}M=M\}.
\eeq
Contrarily, one can define the UCS  coadjoint orbit algebroid of $M$ as
\beq
{\cal O}_M=\{\un\mX\in \text{UCS} \ \vert \ ad^*_{\un\mX}M\neq M \}.
\eeq
The integral lines of these elements of the UCS then define the coadjoint orbits. Although we are here talking about the local tangent plane at the point in UCS$^*$, and confine our attention to the local action, we will colloquially refer to ${\cal O}_M$ as the coadjoint orbit.

As in the case of $\mh^*$, one can here perform a unified treatment of orbits and stabilizers using the KKS symplectic $2$-form:
\beq\label{ucskks2}
\Omega_M: {\cal O}_M\otimes {\cal O}_M\to \RR, \qquad \Omega_M(\un\mX,\un\mY)=\langle M, [\un\mX,\un\mY]\rangle.
\eeq
By definition, the KKS $2$-form is non-degenerate on the orbits. The analysis of the various orbits and different points in UCS$^*$ reduces then to a full characterization of $\Omega$.

There is already a hint that the underlying structure is an algebroid at the algebraic level. Indeed, one observes that the basis $\{\un\pa_\beta,\un t^a{}_b,\un t_a\}$ of the UCS satisfies\footnote{As we will discuss further below, this holds in a local trivialization, so only in a local patch.}
\beqn\label{comm}
\left[\un\pa_\beta,\un \pa_\gamma\right]=\left[\un\pa_\beta,\un t^a{}_b\right]=\left[\un\pa_\beta,\un t_a\right]=0,\quad \left[\un t^a{}_b,\un t^c{}_d\right]
=
\delta^c{}_b\un t^a{}_d -\delta^a{}_d\un t^c{}_b,
\quad
\left[\un t^a{}_b,\un t_c\right]=-\pi^a{}_c{}^d{}_b\un t_d,
\quad
\left[\un t_b,\un t_c\right]=\un 0,
\eeqn
which tells us that the non-trivial $\dS$ action comes entirely from the explicit $\sigma$ dependence of the components $\alpha_\beta$, $\theta^a{}_b$, and $b^a$. From this perspective, $\dS$ plays a special role, and indeed we will associate it to the reparameterization invariance of the base, in the algebroid picture below.

Using \eqref{comm}, the Lie bracket of two elements $\un\mX=\xi^\beta(\sigma)\un\pa_\beta+\theta^a{}_b(\sigma)\un t^b{}_a+ b^a(\sigma)\un t_a$,
 and $\un\mY=\zeta^\beta(\sigma)\un\pa_\beta+\theta'^a{}_b(\sigma)\un t^b{}_a+ b'^a(\sigma)\un t_a$ of the UCS is given by
\beqn
ad_{\un\mX}\un\mY=[\un\mX,\un\mY]=
[\un\xi,\un\zeta]^\beta\un\pa_\beta+\big(\un\xi(\theta'^b{}_a)-\un\zeta(\theta^b{}_a)-[\theta,\theta']^b{}_a\big)\un t^a{}_b+\big(\un\xi(b'^a)-\un\zeta(b^a)-\theta^a{}_bb'{}^b-\theta'{}^a{}_bb^b\big)\un t_a,
\eeqn
which, upon applying the defining representation \eqref{defrep}, is by construction equal to \eqref{LieMsubexpl}. From this we evaluate
\beqn\label{varAlg}
\langle ad^*_{\un\mX}M,\un\mY\rangle=-\langle M,ad_{\un\mX}\un\mY\rangle
&=&\int_S vol_S\Big(-i_{[\un\xi,\un\zeta]}\alpha+J^a{}_b\big(-\un\xi(\theta'^b{}_a)+\un\zeta(\theta^b{}_a)+ [\theta,\theta']^b{}_a
\big)\nonumber\\
&&+ P_a(-\un\xi(b'^a)+\un\zeta(b^a)+\theta^a{}_bb'{}^b-\theta'{}^a{}_bb^b)\Big).
\eeqn
We would like to obtain the variations $\{\delta_{\un\mX}\alpha,\delta_{\un\mX}J^a{}_b,\delta_{\un\mX}P_a\}$ from this expression. Special attention should be devoted to $vol_S$, because it transforms under $\delta_{\un\mX}$ non-trivially. So in order to read the variations one should compare \eqref{varAlg} with\footnote{Perhaps a better notation here would be to write \[ \langle *M,\un\mY\rangle=\int_S *M(\un\mY),\] and then interpret the variation as $\langle \delta_{\un\mX}*M,\un\mX\rangle$. That is, the integral involves the top form dual to $M(\un\mY)$ and we are varying that. In \cite{Donnelly:2020xgu}, a similar analysis was performed, but densities were introduced rather than tensors, to incorporate the $vol_S$ contributions.}
\beqn
\delta_{\un\mX}\langle M,\un\mY\rangle
&=&\int_S \Big(\zeta^\beta\delta_{\un\mX}(\alpha_\beta vol_S)+\theta'^a{}_b\delta_{\un\mX}(J^b{}_a vol_S)+b'^a\delta_{\un\mX}(P_a vol_S)\Big),
\eeqn
where $\zeta^\beta\delta_{\un\mX}(\alpha_\beta vol_S)\equiv (i_{\un\zeta}\delta_{\un\mX}\alpha)vol_S+(i_{\un\zeta}\alpha)\delta_{\un\mX}vol_S$.
Using $\delta_{\un\mX}vol_S=\nabla_\beta\xi^\beta vol_S$, and assuming $S$ has no boundary, we  find the variations 
\beqn\label{UCSdeltaalpha}
\delta_{\un\mX}\alpha&=& {\cal L}_{\un\xi}\alpha+J^a{}_bd\theta^b{}_a+P_a db^a,\\
\delta_{\un\mX}J^a{}_b&=&\un\xi(J^a{}_b)+[J,\theta]^a{}_b  -b^aP_b,
\label{UCSdeltaJ}\\
\delta_{\un\mX}P_a&=&\un\xi(P_a)+P_b\theta^b{}_a.\label{UCSdeltaP}
\eeqn
In this expression, we introduced the exterior derivative $d$ on $S$ and the associated Lie bracket acting on forms as ${\cal L}=id+di$. In addition to the assumption that $S$ is without boundary, we have also assumed smoothness, necessarily required for the manipulations performed here. 

In our previous discussion of the coadjoint orbits of $\mh$, we were able to convert the KKS form from a function of the transformation parameters to a function of the variations of the covector space (see for example eq. \eqref{KKS} versus \eqref{KKSdJdP}). In the present case (including diffeomorphisms) such a process is much more challenging as it would inevitably be non-local on $S$. In the following subsection, we will reinterpret the current construction in terms of a certain Atiyah Lie algebroid. It is perhaps then the case that one could manage the conversion  via the introduction of corresponding groupoids, but this is beyond the scope of the present paper. Nevertheless, we take this as a strong indication that the proper local geometric structure on $S$ is an algebroid over $S$. As we describe in the next subsection, this will allow for a completely geometric account of the representation theory of the ECS and ACS.

Indeed, instead of attempting the aformentioned conversion process, we continue by considering the restriction of the full KKS form
\beqn\label{KKSUCS}
\Omega(\un\mX,\un\mY)=\int_S vol_S\Big(i_{[\un\xi,\un\zeta]}\alpha+J^a{}_b\big(\un\xi(\theta'^b{}_a)-\un\zeta(\theta^b{}_a)- [\theta,\theta']^b{}_a
\big)+ P_a(\un\xi(b'^a)-\un\zeta(b^a)-\theta^a{}_bb'{}^b+\theta'{}^a{}_bb^b)\Big)
\eeqn
to the ECS and ACS subalgebras. By such restrictions, we mean the analogues of the discussion in Section \ref{sec3}. These are restrictions on the $(\alpha,\theta,b)$ in eqs. (\ref{UCSdeltaalpha}--\ref{UCSdeltaP}). For ECS, we obtain
\beqn\label{ECSdeltaalpha}
\delta_{\un\mX_s}\alpha&=& {\cal L}_{\un\xi}\alpha+\bar J^a{}_bd\bar\theta^b{}_a+P_a db^a,\\
\delta_{\un\mX_s}\bar J^a{}_b&=&\un\xi(\bar J^a{}_b)+[\bar J,\bar\theta]^a{}_b  -b^aP_b+\tfrac12 b\cdot P \delta^a{}_b,
\label{ECSdeltaJ}\\
\delta_{\un\mX_s}P_a&=&\un\xi(P_a)+P_b\bar\theta^b{}_a,\label{ECSdeltaP}
\eeqn
while for ACS we have
\beqn\label{ACSdeltaalpha}
\delta_{\un\mX_w}\alpha&=& {\cal L}_{\un\xi}\alpha+\tfrac12{\cal J}dw+P_a db^a,\\
\delta_{\un\mX_w}{\cal J}&=&\un\xi({\cal J})  -b^a P_a,
\label{ACSdeltaJ}\\
\delta_{\un\mX_w}P_a&=&\un\xi(P_a)+\tfrac12 wP_b.\label{ACSdeltaP}
\eeqn
We would like to stress the importance of these equations. They show how any field variation dictated by the ECS and the ACS are realized at points of  UCS$^*$. This is the analogous of the discussion for $\mh$ and its two sub-algebras $\mh_s$ and $\mh_w$ in the previous section. We will now show how these results are accommodated into the general theory of Atiyah Lie algebroids. 

\subsection{Lie Algebroids}\label{sec4.2}

In the previous section we have introduced the invariant pairing
\beq\label{pair}
\langle M,\un\mX\rangle=\int_S vol_S\, M(\un\mX),
\eeq
where 
\beq\label{localmXM}
\un\mX=\xi^\alpha\un\pa_\alpha+\theta^a{}_b\un t^b{}_a+b^a\un t_a,\qquad
M(\un\mX)=i_{\un\xi}\alpha+\theta^a{}_bJ^b{}_a+b^aP_a.
\eeq
We have also seen how introducing $\dS$ complicates the algebraic analysis, and calls for a deeper geometric understanding of the UCS.
In this section, we note that $\un\mX$ can be interpreted as a section of an Atiyah Lie algebroid \cite{Atiyah:1957,Atiyah:1979iu} over $S$, associated to the group $H=GL(2,\RR)\ltimes \RR^2$, in terms of which eqs. \eqref{localmXM} can be interpreted as formulas valid within a local trivialization. The construction is as follows (details can be found in our recent exploratory paper \cite{Ciambelli:2021ujl}, from which most conventions are taken). We first introduce the principal $H$-bundle $\pi:P_c\to S$. The corresponding Atiyah Lie algebroid is obtained as the quotient of $TP_c$ by the right action of $H$, $A_c=TP_c/H$. This is a vector bundle of rank $n+6$ over $S$ ($n=\dim S$) that is equipped with an anchor map $\rho:A_c\to TS$ whose kernel is the vertical sub-bundle of $A_c$ isomorphic to the adjoint bundle $L_c=P_c\times_{Ad_H} \mh$. That is, there is a short exact sequence
\beq\label{shortExactSeq}
\begin{tikzcd}
0
\arrow{r} 
& 
L_c
\arrow{r}{j} 
& 
A_c
\arrow{r}{\rho} 
& 
TS
\arrow{r} 
&
0
\end{tikzcd}.
\eeq
Locally, we can think of $L_c$ as having fibres $\mh$, and thus the value of a section of $L_c$ at a point $p\in U\subset S$ is an element of the Lie algebra $\un\mu_p\in\mh$. A local trivialization of $A_c$ on $U\subset S$  is a map $\tau:A_c|_U\to TU\times L_c|_U$, and so we can always think of a section $\un\mX$ of $A_c$ as given locally by a vector on $S$ together with an element of the Lie algebra, that is 
\beq
\tau(\un\mX)=\xi^\alpha\un\pa_\alpha+\mu^A\un t_A,
\eeq
which explicitly associates a section of $A_c$ with an element of the UCS, that is $\un\mX_p \to (\un\xi_p,\un\mu_p)$. Here, we reiterate that $\sigma^\alpha, (\alpha=1,\dots,n)$ are  local coordinates on $U$ and we introduced $\un t_A, (A=1,\dots,6)$, a local frame for $L_c$.

Each of the vector bundles described above is equipped with a skew bracket: $L_c$ has the Lie algebra bracket $[\cdot,\cdot]_L$, $TS$ the Lie bracket of vector fields $[\cdot,\cdot]$, while the bracket $[\cdot,\cdot]_A$ on $A_c$ is such that $\rho$ and $j$ are morphisms, that is, given two (local) sections $\un\mX$ and $\un\mY$ of $A_c$ and two sections $\un\mu,\un\nu$ of $L_c$, we have
\beq
\rho([\un\mX,\un\mY]_A)=[\rho(\un\mX),\rho(\un\mY)],\qquad
j([\un\mu,\un\nu]_L)=[j(\un\mu),j(\un\nu)]_A.
\eeq
A connection on the algebroid $A_c$ is given by a map $\sigma:TS\to A_c$ (not to be confused with the local coordinates $\sigma^\alpha$ on $S$) and $\omega:A_c\to L_c$ with $\omega\circ\sigma=0$. The curvature of the connection is a measure of the failure of $\sigma$ and $-\omega$ to be morphisms of the brackets; we again refer to \cite{Ciambelli:2021ujl} for more details. 
The map $\sigma$ can be interpreted as an Ehresmann connection, which provides a lift of a vector field in $TS$ to a section of the horizontal sub-bundle $H_c$ of $A_c$, such that $A_c=H_c\oplus V_c$. In a local trivialization, we can write a `split' basis of sections of $H_c$, denoted $\un E_{\un\alpha}$ with $({\un \alpha}=1,\dots,n)$, and of sections of $V_c$, denoted $\un E_{\un A}$ with $({\un A}=1,\dots,6)$, as
\beq\label{trivbasisA}
\tau(\un E_{\un\alpha})=\tau(\sigma)^\alpha{}_{\un\alpha} (\un\pa_\alpha+ a^A_\alpha(\sigma) \un t_A),\qquad
\tau(\un E_{\un A})=\tau(\sigma)^A{}_{\un A} \un t_A,
\eeq
respectively. Note that here we are using a short form notation $\un t_A$, but in the case at hand, we could also use the notation $(\un t^a{}_b,\un t_a)$. Similarly, we have written the connection coefficients here as $a^A_{\alpha}$ but if we again regard the $\un t_A$ as the pair $(\un t^a{}_b,\un t_a)$, we have a pair of gauge fields $(a_{\alpha}^{(0)}{}^a,a_{\alpha}^{(1)}{}^a{}_b)$. These form an $H$-connection, in the sense that on an overlap $U_i\cap U_j$, we have
\beq
(a_{i,\alpha}^{(0)}{}^a,a_{i,\alpha}^{(1)}{}^a{}_b)=J^\beta{}_\alpha 
(R^{-1})^a{}_c (a_{j,\beta}^{(0)}{}^c+\pa_\beta b^c-a_{j,\beta}^{(1)}{}^c{}_db^d,
-\pa_\beta R^c{}_b+a_{j,\beta}^{(1)}{}^c{}_d R^d{}_b).
\eeq

Furthermore, what we previously called $M\in$ UCS$^*$ can now be interpreted as a section of the dual bundle $A_c^*$. As usual, a section of the dual bundle, $M\in\Gamma(A_c^*)$ is defined as a map $M:A_c\to C^\infty(S)$. Thus, 
we can import all the results of Subsection \ref{sec4.1}, and re-interpret them on this Atiyah Lie algebroid.
The basis $\{E^{\un\alpha},E^{\un A}\}$ for $A_c^*$ that is dual to \eqref{trivbasisA} is given in a local trivialization by\footnote{In Ref. \cite{Ciambelli:2021ujl}, we denoted the inverse map $\tau^*:A_c^*|_U\to T^*U\oplus L_c^*|_U$. To avoid confusion with the algebroid pullback, we now refer to this map as $\bar\tau$ and denote its matrix elements by $\tau^{-1}$.
Also as described in \cite{Ciambelli:2021ujl}, we enforce $\tau\circ j = Id_{L|_U}$. }
\beq\label{trivbasisAdual}
\bar\tau(E^{\un\alpha})=(\tau^{-1}(\sigma))^{\un\alpha}{}_\alpha d\sigma^\alpha,\qquad
\bar\tau(E^{\un A})=(\tau^{-1}(\sigma))^{\un A}{}_A(t^A-a^A_\alpha(\sigma)d\sigma^\alpha).
\eeq

The coadjoint orbit analysis discussed previously can now be formulated more geometrically in terms of $A_c^*$, rather than UCS$^*$. The purpose of the orbit analysis is to better understand  representations. 
The algebroid geometric structure is well-suited to do so, for any representation of $\mh$ corresponds to an associated  bundle $E\to S$ whose fibres correspond to the representation space. Furthermore, one can construct a notion of differentiation of sections of $E$ by establishing a morphism between $A_c$ and another Lie algebroid, the algebroid of derivations of sections of $E$, called $Der(E)$, such that 
\beq\label{DerEshortExactSeqGen}
\begin{tikzcd}
& 
L_c
\arrow{r}{j} 
\arrow{dd}{v_E}
& 
A_c
\arrow{dr}{\rho} 
\arrow{dd}{\phi_E} 
& 
&
\\
0
\arrow{ur} 
\arrow{dr} 
&&&
TS
\arrow{r} 
&
0
\\
& 
End(E)
\arrow{r}{j_E} 
& 
Der(E)
\arrow{ur}{\rho_E} 
& 
&
\end{tikzcd}.
\eeq
For $\psi\in\Gamma(E)$ a local section of $E$, one then has $\phi_E(\un\mX)(\psi)=\hatd\psi(\un\mX)$, with $\hatd\psi\in\Gamma(A_c^*\times E)$. This can be extended to sections of $\wedge^kA_c^*\times E$, with $\hatd:\wedge^kA_c^*\times E\to \wedge^{k+1}A_c^*\times E$. Given a connection on $A_c$, there is an induced connection on $Der(E)$ such that the horizontal part of $\hatd\psi$ can be interpreted as a covariant derivative, $\hatd\psi(\un\mX_H)=\nabla^E_{\rho(\un\mX)}\psi$.  In the diagram above, the $End(E)$ can be thought of as giving rise to a matrix representation of $\mh$. 

Let us explore possible representations in our current context.
In most applications, the associated bundles are taken to be vector bundles, whose transition functions are linear maps. 
In the case at hand, where $H=GL(2,\RR)\ltimes \RR^2$, there is an important associated bundle $\pi_\aff:\aff\to S$ which is a rank-2 {\it affine} bundle. This bundle is said to be modeled on a rank-2 vector bundle (with fibre $V_\aff$), with transition functions given by affine maps. That is for $U_i,U_j\subset S$, a local section $\psi\in\Gamma(\aff)$ satisfies (a,b=1,2)
\beq\label{transfnB}
(\sigma_i^\alpha,\psi^a_i)=(\sigma_i^\alpha(\sigma_j),R_{ij}{}^a{}_b\psi^b_j+b_{ij}{}^a),
\eeq
on $U_i\cap U_j$, 
where we are writing the components of the section with respect to a basis for $V_\aff$. Thus the transition functions are determined by $(R_{ij}{}^a{}_b, b_{ij}{}^a)$, which indeed correspond to an element of $H$. Infinitesimally, we have
\beq\label{transfnBinf}
(\sigma_i^\alpha,\psi^a_i(\sigma_i))=(\sigma_j^\alpha-\xi^\alpha(\sigma_j),{\cal L}_{\un\xi}\psi^a_j(\sigma_j)+\theta^a{}_b(\sigma_j)\psi^b_j(\sigma_j)+b^a(\sigma_j)),
\eeq
so we see that the infinitesimal transformation of $\psi^a$ is given by a local element of the UCS.  The affine bundle might be referred to as the fundamental representation.
Note that if we had called the section $u\in\Gamma(\aff)$, we could regard the components $u^a$ as local fibre coordinates, and then \eqref{transfnB} appear as diffeomorphisms of the total space of $\aff$, restricted to lie in $H$. We will argue later that there is a `semi-classical' correspondence between the bundle $\aff$ and a  spacetime $M$, near the corner, in the sense that the restricted diffeomorphisms are precisely those that have non-zero charges in a classical theory. This is the reason why we focus on this particular representation in the rest of this subsection.

The corresponding structure for the dual bundles is given by 
\beq\label{DerEshortExactSeqGen}
\begin{tikzcd}
& 
L_c^*
\arrow{dl}
& 
A_c^*
\arrow{l}{j^*} 
& 
&
\\
0
&&&
T^*S
\arrow{dl}{\rho_\aff^*} 
\arrow{ul}{\rho^*} 
&
0
\arrow{l} 
\\
& 
End(\aff)^*
\arrow{uu}{v_\aff^*}
\arrow{ul}
& 
Der(\aff)^*
\arrow{l}{j_\aff^*} 
\arrow{uu}{\phi_\aff^*} 
& 
&
\end{tikzcd}.\label{derBdiagram}
\eeq

Consider a section $M$ of $A_c^*$. In a local split basis, we can write $M=M_{\un \alpha}E^{\un\alpha}+M_{\un A}E^{\un A}$. Given the above discussion, in a local trivialization, one then has
\beqn\label{tauM}
\bar\tau (M)&=&M_{\un A}(\tau^{-1})^{\un A}{}_A(t^A-a^A_\alpha d\sigma^\alpha)+M_{\un\alpha} (\tau^{-1})^{\un\alpha}{}_\alpha d\sigma^\alpha\nonumber\\
&\equiv &
M_{ A}(t^A-a^A_\alpha d\sigma^\alpha)+M_{\alpha} d\sigma^\alpha.\nonumber\\
&=&M^b{}_at^a{}_b+M_at^a
+(M_{\alpha}-a^{(0)}_\alpha{}^aM_a-a^{(1)}_\alpha{}^a{}_bM^b{}_a) d\sigma^\alpha.
\eeqn
The first three terms in the last line of this expression relate to $GL(2,\RR)$, $\RR^2$ and $Diff(S)$ respectively and they coincide in form with the Noether charge aspects found on a finite distance corner in a classical spacetime \cite{Ciambelli:2021vnn}, we will expand on this in the next subsection. 
The last term in \eqref{tauM} deserves some comment. If we compare this result to that found in the Einstein-Hilbert theory on a classical spacetime \cite{Ciambelli:2021vnn}, in the latter case, only the third term, involving $a^{(1)}$, appeared. This can be explained first by the fact that the embedding condition implied that $a^{(0)}$ pulls back to zero on $S$, while the $Diff(S)$ that appeared there was actually associated with changes of coordinates for the embedding $\phi(S)$ as opposed to changes of the intrinsic coordinates on $S$ (which would be pure gauge from the point of view of the bulk theory defined on the spacetime manifold). 
We can then re-interpret these restrictions on the section of $A^*_c$ as a sort of 'gauge fixing' on the various fundamental fields.

We would like to rewrite \eqref{tauM} in terms of the associated bundle $\aff$. Denoting the corresponding section of $Der(\aff)^*$ by $M_\aff$, we have $\phi_\aff^*(M_\aff)=M$.
 Now, given the local trivialization $\bar\tau$ of $A_c^*$, there is a corresponding trivialization $\bar\tau_\aff:Der(\aff)^*|_U\to T^*U\oplus End(\aff)^*|_U$, satisfying $\bar\tau\circ\phi_\aff^*=(1\otimes v_\aff^*)\circ\bar\tau_\aff$. 
We can take a basis for $Der(\aff)$ as $(\un F_{\un{\alpha}},\un F_{\un{A}})=(\phi_\aff(\un E_{\un\alpha}),\phi_\aff(\un E_{\un A}))$,
and the dual basis for $Der(\aff)^*$ is then $(F^{\un\alpha},F^{\un A})$. We thus obtain
\beq
\bar\tau_\aff(F^{\un\alpha})=(\tau^{-1}(\sigma))^{\un\alpha}{}_\alpha d\sigma^\alpha,\qquad
\bar\tau_\aff(F^{\un A})=(\tau^{-1}(\sigma))^{\un A}{}_A(v^A-a^A_\alpha(\sigma)d\sigma^\alpha),
\eeq
where the upper triangular matrices $v^A$ are a basis for $End(\aff)^*$, such that $v_\aff^*(v^A)=t^A$, with $v^A(v_\aff(\un t_C))=\delta^A{}_C$. The latter can be interpreted as a matrix trace. 

Then, we can eventually write \eqref{tauM} from the point of view of  $\aff$ as
\beqn
\bar\tau_\aff (M_\aff)&=&
M_\aff{}_{\un\alpha}(\tau^{-1}(\sigma))^{\un\alpha}{}_\alpha d\sigma^\alpha
+M_\aff{}_{\un A}(\tau^{-1}(\sigma))^{\un A}{}_A(v^A-a^A_\alpha(\sigma)d\sigma^\alpha)\nonumber\\
&\equiv&
M_\aff{}_{\alpha} d\sigma^\alpha
+M_\aff{}_{A}(v^A-a^A_\alpha(\sigma)d\sigma^\alpha)
\nonumber\\
&=&M_\aff{}^b{}_av^a{}_b+M_\aff{}_av^a
+(M_\aff{}_{\alpha}-a^{(0)}_\alpha{}^aM_\aff{}_a-a^{(1)}_\alpha{}^a{}_bM_\aff{}^b{}_a) d\sigma^\alpha.\label{tauM1}
\eeqn
This result is the starting point to make contact with the classical spacetime representation. Indeed, if we start with $A_c^*$ alone, there is no notion of normal coordinates and classical spacetime reconstruction. it is only via the affine bundle $\aff$ that these quantities appear. In our endeavor to understand gravity better, $\aff$ is the basic ingredient to probe classical aspects and representations of the UCS.

Since this subsection contains various new perspectives on the corner proposal, we would like to offer a brief summary before continuing. The rewriting of the UCS in terms of an Atiyah Lie algebroid over $S$ sheds light on its geometric aspects, and allows the inclusion of $\dS$ in a natural way, as the reparameterization invariance of the base manifold. Clearly, the issue faced in the previous section in trying to invert the KKS 2-form still persists, and a full classification of the UCS representations is for now out of reach. Nonetheless, we now have a different way to appreciate certain representations of the UCS, via the construction of associated bundles to the algebroid. We have in particular identified one representation, on an affine bundle modeled on a rank-$2$ vector bundle, in which $2$ 'normal' directions naturally appear on the fibres. We have moreover seen that the components of a section $M$ of $A^*_c$, or the corresponding section $M_B$ of $Der(\aff)^*$, have the same structure as the fields that enter into the non-zero charges of a finite distance corner on a classical gravitational  spacetime. This will become more precise in the next section when we discuss moment maps, which relate classical on-shell field spaces to specific orbits inside $A_c^*$. So there are two  main results in this subsection. The first is the construction of $A^*_c$, which contains the image of both finite distance and asymptotic corners moment maps. The second is the identification of $\aff$ as the first step in the reconstruction of a classical spacetime, if the starting point is $A^*_c$. In the corner proposal, we  in fact stipulate that corners are the atomic constituents of gravity. We advocate that this is the correct path toward a better understanding of gravity. In particular, other representations of $A^*_c$ might be  relevant, and could instruct us about quantum properties of the theory.

\subsection{Moment Map}\label{sec4.3}

So far, we have extrapolated the UCS from classical gravity and then worked only intrinsically to the corner. We now connect the classical field space to the dual algebroid $A^*_c$ via the moment map. Before doing so in specific examples, we offer the general analysis. 

Calling $\chi$ the set of physical classical fields of a dynamical theory, the moment map is the map $\mu:\chi\to A^*_c$ that links these fields to a point in $A^*_c$. Then, the symmetry algebra acts on the tangent space at the point and infinitesimally moves it on the  orbit via the coadjoint action \eqref{adad}. The orbit is a symplectic manifold whose non-degenerate $2$-form is precisely the KKS $2$-form  \eqref{ucskks2}.

The moment map applies only to symplectomorphisms, and therefore, it is properly defined only for integrable Noether charges. That is, calling $H_{\un\mX}$ the Noether charge of a given dynamical theory with symplectic $2$-form $\Omega$, we have
\beq\label{int}
\delta H_{\un\mX}=-I_{V_{\un\mX}}\Omega,
\eeq
where $V_{\un\mX}$ is the symplectomorphism associated to the symmetry generator $\un\mX$ and $I$ and $\delta$ are the interior product and exterior derivative on $\chi$, respectively. Our extended phase space of gravity proposed in \cite{Ciambelli:2021vnn,Ciambelli:2021nmv} (see also \cite{Freidel:2021dxw}) is such that all gravitational diffeomorphism surface charges are integrable, and thus the whole field space can be mapped to the dual algebroid $A_c^*$. We consider this a very important improvement, whose consequences are yet to be fully explored. We will return to this discussion momentarily. 

In practice, the identification of the moment map is performed via the pairing introduced in \eqref{ucspair}. Using the notation $\mu(\chi)=M$, one has that
\beq\label{mupair}
\langle\mu(\chi),\un \mX\rangle=\int_S vol_S\, \mu(\chi)(\un\mX)\equiv \mu(H_{\un\mX}),
\eeq
where in the last equality we introduced a compact, yet slightly abusive, notation. Then, using this notation and requiring that the moment map relates the symplectic $2$-form $\Omega$ on $\chi$ to the KKS $2$-form $\Omega_{\mu(\chi)}$ at the point $\mu(\chi)\in A^*_c$, that is,\footnote{In this equation and henceforth, we are  using the notation $\delta_{\un\mX}$ to denote the field space variation along $V_{\un\mX}$, that is, the field space Lie derivative $I_{V_{\un\mX}}\delta+\delta I_{V_{\un\mX}}$. This should not be confused with our previous use of the symbol in Section \ref{sec4.1}. }
\beq
\Omega_{\mu(\chi)}(\un\mX,\un\mY)=-\langle ad^*_{\un\mX}(\mu(\chi)),\un\mY)\rangle 
= \mu(H_{[\un\mX,\un\mY]_A})
=-\mu(\delta_{\un\mX}H_{\un\mY})=\mu(I_{V_{\un\mX}}I_{V_{\un\mY}}\Omega),
\eeq
one obtains that $\mu$ is compatible with the field space variation $\delta_{\un\mX}$ on $\chi$:
\beq
 \mu\circ \delta_{\un\mX}=ad^*_{\un\mX}\circ \mu.
\eeq
This means that the charge algebra is realized on the orbit in $A^*_c$ via the coadjoint action.

This discussion has followed the typical path from a classical theory to the abstract dual algebra and, indeed, there was no need to introduce $\aff$. The theme of this work is to make one further step  and, once $A^*_c$ is understood, take it as the starting point and study its representations. If we knew little or nothing about the classical field space (which is the state of the art concerning quantum representations and observables), the first step would have been to introduce an associated bundle on which one can identify two fibre coordinates that play the role of the 'normal directions' in the vicinity of a corner in a classical spacetime. Then, the algebra acting at points in $A^*_c$ is mapped to derivations of this bundle, which can be packaged and reinterpreted as probing the normal directions. This is exactly why $\aff$ is a fundamental ingredient when thinking of a classical field space from the intrinsic algebroid viewpoint. It is this corner-to-bulk shift of paradigm that allowed us to appreciate the ECS and ACS as sub-algebras of the UCS. Although a complete reconstruction picture of a classical spacetime  is still under investigation, we have at least set the stage here  to do so. This is also a framework where quantum features of gravity seem within reach, as they could be revealed studying other representations and associated bundles of the algebroid $A_c^*$, without any interference from classical inputs, such as dynamics and, even more importantly, the notion of a metric and spacetime.

\subsection{ECS and ACS Moment Maps}\label{sec4.4}

In this concluding subsection, we explicitly construct the moment maps of finite distance and asymptotic corners, since we have introduced all the important ingredients. The common thread of this manuscript is how to unify the treatment of ECS and ACS inside the UCS, the universal symmetry algebra. This applies also to the moment maps: while the intrinsic moment maps for finite distance and asymptotic corners link their solution spaces respectively to points on the dual algebras ECS$^*$ and ACS$^*$, we have seen how to directly interpret them on the dual algebra UCS$^*$. 

\paragraph{Finite-distance corners:} Following the conventions established in \cite{Ciambelli:2021vnn}, we begin with the bulk metric parameterization of an arbitrary spacetime adapted to a corner
\beq\label{randers}
g=
h_{ab}(u,x)n^a\otimes n^b+\gamma_{ij}(u,x)dx^i\otimes dx^j,
\eeq
and expand the metric constituents order by order as in (\ref{exp1},\ref{exp2}). This parameterization is well-suited and preferred, since the metric constituents transform as affine tensors under the full UCS (see \cite{Ciambelli:2021vnn} for more details). The Noether charges are given by 
\beq
H_{\un\mX}
=\int_S vol_S\ \Big(\xi_{(1)}{}^a{}_bN^b{}_a+\xi_{(0)}^jb_j+\xi_{(0)}^ap_a\Big)\label{fullcharge},
\eeq
with\footnote{Note that the first and last equations can be rewritten as
\beqn
N^a{}_b=-\frac{1}{\sqrt{-\det h^{(0)}}}\varepsilon^{ac}h^{(0)}_{cb},\qquad
p_a=-\frac{1}{\sqrt{-\det h^{(0)}}}\tfrac12\varepsilon^{bc}(h^{(1)}_{ab,c}-h^{(1)}_{ac,b}).
\eeqn
}
\beqn\label{N}
N^b{}_a =
\sqrt{-\det h^{(0)}}
 h_{(0)}^{bc}\varepsilon_{ca},\qquad
b_j = -N^b{}_aa^{(1)}_j{}^a{}_b,\qquad
p_d = \tfrac12 N^{a}{}_{c}h_{(0)}^{cb}(h^{(1)}_{db,a}-h^{(1)}_{da,b}).
\eeqn

Clearly the classical result is of the same form as written above, in eq. \eqref{ucspair} for example. We note however that $N^a{}_b$ satisfies two conditions: first, it is traceless, which should be regarded as the reduction of the symmetry to ECS for such finite distance corners. Second, it satisfies $\tfrac12tr\, N^2=1$ (equivalently, $\det N=-1$); this should be regarded as a special property of the point on the ECS orbit that we happen to find ourselves. This statement can be understood in two ways: first, $tr\, N^2$ is not a Casimir, and so would vary along an orbit; second, and equivalently, we have made a choice by writing the metric in the form \eqref{randers} -- if it were modified by $g\to g/\Omega(u,x)^2$ then the net effect on the charges is to take $N^a{}_b\to N^a{}_b/\Omega^{(0)}(x)^n$. Note that the assumption that $\Omega^{(0)}(x)\neq 0$ is equivalent to the statement that the corner is at finite distance.

Now let us go back to the algebroid construction. For a finite distance corner, given that the symmetry  reduces to ECS$^*$, we have an invariant tensor $\varepsilon^{ab}$, and using it, we can convert a section of $End(\aff_s)^*$ to a symmetric bilinear form
\beq
\varepsilon: End(\aff_s)^*\to S^2\aff_s.
\eeq
Here by $\aff_s$, we mean the affine bundle associated to the G-structure $A_s$ whose structure group is reduced to $H_s$ (operationally, this just means that $M^a{}_b$ is traceless).
In particular, a section of $S^2\aff_s$ is a pair $(h_{(0)}^{ab},h_{(1)}^a)$ where $h_{(0)}$ is symmetric, and we identify\footnote{As a section of $S^2\aff$, the pair $(h_{(0)}^{ab},h_{(1)}^a)$ can be organized into a matrix $\begin{pmatrix} h_{(0)}^{ab} & h_{(1)}^a\cr h_{(1)}^b&1\end{pmatrix}$ and transform under $\mh_s$ as 
\beq
\delta_{\un\mu}h_{(0)}^{ab}=
 \bar\theta^a{}_ch_{(0)}^{cb}+\bar\theta^b{}_dh_{(0)}^{ad}+b^ah_{(1)}^{b}+h_{(1)}^{a}b^b,\qquad
 \delta_{\un\mu}h_{(1)}^a=
\bar\theta^a{}_ch_{(1)}^{c}+b^a.
\eeq
}
\beqn\label{N}
h_{(0)}^{ab}:=M^a{}_c\varepsilon^{cb},\qquad
h_{(1)}^{a}:=M_b\varepsilon^{ba}.
\eeqn
By inverting $h_{(0)}$, and using it to lower indices, we can equivalently regard this data as a section $(h^{(0)}_{ab},h^{(1)}_a)$ of $S^2\aff_s^*$, and we identify
\beq
h^{(1)}_a=\tfrac12\varepsilon^{bc}(h^{(1)}_{ab,c}-h^{(1)}_{ac,b}).
\eeq
We note then that this data, along with the connection on the affine bundle and a metric on $S$, is equivalent to the data contained in a bulk metric near a corner, in the sense that these are the components of the bulk metric that contribute to the Noether charges for a finite distance corner. To recap then, we have seen that the affine bundle $\aff$ may be regarded as a local model for a bulk spacetime, while a section of $Der(\aff)^*$ can be interpreted as giving rise to a local metric on that bulk spacetime.

We can interpret \eqref{N} as essentially a moment map: we regard the field space as 
$\chi=\{h^{(0)}_{ab},a^{(1)}_j{}^a{}_b,h^{(1)}_{a[b,c]}\}$, and the moment map as given by
\beq
\mu_{s}(N^a{}_b)=J^a{}_b\qquad \mu_{s}(p_a)=P_a\qquad \mu_{s}(b_i)=\alpha_i.
\eeq
More precisely, this may be regarded as a ``pre-moment map" that maps onto ECS$^*$, but not obviously to the non-degenerate orbits.
However, of the five $N^a{}_b,p_a$, only four will be independent on an orbit.\footnote{Indeed one finds that evaluating the charges on a finite distance corner in any particular classical geometry, only a subset of the $\mh_s$ charges are independent, which we now understand as this restriction to the $\mh_s$-orbit.} To demonstrate this, we can construct the Casimir
\beq
{\cal C}_3=\int_S vol_S\ p_aN^a{}_b\varepsilon^{bc}p_c,
\eeq
and show that
\beq
ad^*_{\un\mX}{\cal C}_3=0.
\eeq
From a classical bulk perspective, this result would be interpreted as a constraint equation. 

\paragraph{Asymptotic corners:} The corresponding analysis for the asymptotic corner symmetry is more involved on the classical side because of the need for holographic renormalization, and some details will depend on precisely which asymptotic setting (e.g., asymptotically flat or AdS) is studied. Nevertheless, we expect that a moment map relates the renormalized charges to the expressions involving ACS$^*$. At a point on $S$ the ACS reduces to $\mh_w$. The latter is a three-dimensional ideal of $\mh$, whose orbits are at best two-dimensional, because of the existence of Casimirs. 

The pairing \eqref{pair} for ACS is given by
\beq\label{acspair}
\langle M,\un\mX\rangle=\int_S vol_S\, (i_{\un\xi}\alpha+\tfrac12 {\cal J}w+b^aP_a).
\eeq
We stress that the ACS pertains to both asymptotically flat and AdS solution spaces, and so \eqref{acspair} is the starting point to construct both moment maps.

We begin by discussing asymptotically flat spacetimes, using results from \cite{Freidel:2021fxf}. The setup is the following. We work in $4$ spacetime dimensions, the normal coordinates to the corner are $(u,r)$ while coordinates on the corner are $x^A$, and the asymptotic boundary is located at $r\to \infty$. The leading order of the vector field generating asymptotic symmetries is
\beq\label{BMSW}
\un\xi=T(x)\un\pa_u+ Y^A(x)\un\pa_A+W(x)(u\un\pa_u-r\un\pa_r),
\eeq
and generates the so-called Weyl BMS symmetry, \cite{Freidel:2021fxf}. To compare with our setting, where the normal coordinates are expanded around the corner as $u^a\to 0$, we need to perform the change of coordinate $r={1\over \rho}$, such that the conformal boundary is at $\rho\to 0$. Then, eq. \eqref{BMSW} becomes
\beq\label{BMSWrho}
\un\xi=T(x)\un\pa_u+ Y^A(x)\un\pa_A+W(x)(u\un\pa_u+\rho\un\pa_\rho).
\eeq
Comparing with the defining representation introduced in Section \ref{sec2}, this vector field is included in the ACS, with the normal translation along $\un\pa_\rho$ turned off, i.e., $b^\rho=0$. Using the conventions of the present paper, the other parameters are identified  as $T(x)=b^u(\sigma)$, $W(x)=2w(\sigma)$, and $Y^A(x)=\xi^\alpha(\sigma)\delta^A_\alpha$.

The renormalized charges $Q_R$ (referred to as $H_{\un\mX}$ in Section \ref{sec4.3}) in $4$-dimensional Einstein gravity with vanishing cosmological constant, evaluated at the corner on the asymptotic boundary at $u=0$, were found in \cite{Freidel:2021fxf} to be given by
\beq
Q^R=\int_S vol_S \left(T (M-\tfrac12 \bar D_A U^A)+4 W \bar \beta+Y^A(\bar P_A+2\pa_A \bar\beta)\right),
\eeq
where $M,U^A,\bar\beta$, and $\bar P_A$ are pieces of the bulk metric in Bondi gauge, and $\bar D_A$ is the corner covariant derivative. 
The asymptotic solution space is thus parameterized by $\chi=\{M,U^A,\bar\beta,\bar P_A\}$ and, comparing with \eqref{acspair}, we identify the moment map:
\beq
\mu_w(M-\tfrac12 \bar D_AU^A)=P_u,\qquad \mu_w(16\bar\beta)={\cal J},\qquad \mu_w(\bar P_A+2\pa_A\bar\beta)=\delta_A^\beta \alpha_\beta,
\eeq
where in the last expression we converted the corner coordinates $x^A$ into $\sigma^\beta$. Contrary to the finite distance discussion, this is a moment map, that is, it maps directly into the non-degenerate orbits. Indeed, the solution space is here fully on-shell of the bulk equations of motion. From the point of view of $Der(B)$, we observe that the ACS collapses the $4$-dimensional non-affine part of the symmetries to the Weyl singlet. This implies that only the conformal factor of the full boundary metric matters in the solution space. However, in order to understand this better, one should perform an analysis of asymptotic corners with an adapted parameterization (not a gauge) and expansion, as done for finite distance corners in \eqref{randers}. Such an investigation  would be certainly interesting to pursue.

We would like now to turn our attention to asymptotically AdS spacetimes. Our starting point is \cite{Alessio:2020ioh}, so the context is $3$-dimensional Einstein gravity with negative cosmological constant. Working in the Fefferman-Graham gauge with $\rho$ radial direction (conformal boundary at $\rho=0$) and $x^a=(t,\phi)$ boundary coordinates, the leading order of the residual vector field is
\beq
\un\xi=\big(\tfrac12 \bar D_aY^a-\omega(x)\big)\rho\un\pa_\rho+Y^a(x)\un\pa_a,
\eeq
where $\bar D_a$ is the boundary covariant derivative.
This vector field generates a priori a bigger algebra than the ACS, but, once expanded around the asymptotic corner at $(t,\rho)=0$, only $\omega^{(0)}(\phi),Y^t_{(0)}(\phi)$, and $Y^\phi_{(0)}(\phi)$ contribute to the charges, and generate Weyl, supertranslations, and diffeomorphisms of the $1$-dimensional corner, respectively. Thus, the physical asymptotic symmetry algebra around a corner is just $(\dS\loplus \RR)\loplus \RR$, which is again included in the ACS, as advertised. 

The renormalized charges, as computed in \cite{Alessio:2020ioh} in the chiral splitting and expanded around $t=0$, are given by\footnote{Conventions when comparing with \cite{Alessio:2020ioh}: $\Xi_t={\Xi_{++}+\Xi_{--}\over \ell}$, $\Xi_\phi=\Xi_{++}-\Xi_{--}$.}
\beq
Q_{(\omega,Y)}={\ell\over 8\pi G}\int_0^{2\pi}d\phi \Big(Y^t_{(0)}(\phi)\Xi_{t}+Y^\phi_{(0)}(\phi)\Xi_{\phi}-2\ell\omega^{(0)}(\phi)\pa_t\varphi\Big),
\eeq
where $\ell$ is the AdS radius, $G$ is the Newton constant, $\Xi_t$ and $\Xi_\phi$ are the bulk metric field composing the boundary stress tensor, and $\varphi$ is the conformal factor of the boundary metric. These three last quantities compose thus the sourced solution space of the bulk theory, $\chi=\{\Xi_t,\Xi_\phi,\varphi\}$. Then, comparing with \eqref{acspair}, we can identify the moment map:
\beq
\mu_w^{\text{AdS}}(e^{-2\varphi}\Xi_t)=P_u, \qquad \mu_w^{\text{AdS}}(-2\ell e^{-2\varphi}\pa_t \varphi)={\cal J}, \qquad \mu_w^{\text{AdS}}(e^{-2\varphi}\Xi_\phi)=\alpha_\phi.
\eeq
This final result concludes this section. It would be interesting to explore higher dimensions, using the existing literature \cite{Fiorucci:2020xto}, but utilizing the more suitable Weyl-Fefferman-Graham gauge introduced in \cite{Ciambelli:2019bzz}. We leave this analysis for future investigations.

All the results derived previously in this manuscript, on how to simultaneously realize the ECS and ACS orbits inside the UCS, apply here and, in particular, this demonstrates that finite distance and asymptotic corner physics can be described as specific reductions of the UCS. While this analysis has taken a step further in the understanding and unification of the treatment of symmetries of gravity, there are still many open questions that we discuss in the next section.

\section{Outlook}\label{sec5}

In this paper, we have started to explore a new avenue of investigation in which  a minimal amount of symmetry information has been extracted  that we expect to hold in the quantum regime. The identification of the universal symmetry algebra at corners is, in this regard, a lamppost. We thus studied its representations using the orbit method. We have shown that the classical gravitational phase space both at finite distance and asymptotic corners is sent through the moment map to the dual algebra of the UCS, proving that the latter contains and describes both simultaneously. We have furthermore  geometrically realized the UCS as the symmetry group of an Atiyah Lie algebroid associated to a $GL(2,\RR)\ltimes \RR^2$-principal bundle over the corner. This provides the correct arena to study representations of the UCS, each of which corresponds to a specific bundle associated to the Lie algebroid. We have identified the representation giving rise to the asymptotic phase space of a classical spacetime near a corner. 

While our results fit perfectly the finite distance corner analysis, the asymptotic corner should be studied more carefully already at the classical level, and carefully related to the asymptotic limit of finite-distance results. There are two main reasons why asymptotic corners are more complicated to study. The first is that an analysis similar to \cite{Ciambelli:2021vnn} has yet to be done. The latter is a difficult task, due to the second reason, which is holographic renormalization. While tremendous effort has been focussed in recent years on this topic, a full understanding of this procedure for asymptotically flat spacetimes, in particular concerning the consequences it has on the coadjoint orbit analysis is yet to be explored. We plan to return to this in the near future.

In addition, we plan to study the classification of UCS representations, with two main objectives. The first one is to store all the classical information of a spacetime into the dual algebra UCS$^*$, and concretely retrieve various known gravitational solutions as a set of discrete corners with certain field content. We expect this to be possible thanks to the extended phase space introduced in \cite{Ciambelli:2021nmv,Freidel:2021dxw}, whose fundamental role has been already appreciated in the present manuscript. In this regard, it would be rewarding to appreciate the properties of edge modes in a concrete setup, such as corners sitting on a black hole horizon. 

The second, more far-reaching, objective is to understand how certain representations are better suited for a quantum description of the gravitational field content. We expect that the construction of a quantum theory along these lines will lead to unitarity and locality. The presence of the BRST symmetry on Atiyah Lie algebroids \cite{Ciambelli:2021ujl} suggests that some quantum features of gravity are already present in our description, and it is worth investigating. Other questions are expected to find a natural answer in this direction. For instance, the role of gravitational electric-magnetic duality at corners, or S-matrix scattering properties, are questions on our agenda. 

Finally, our recent works, starting with \cite{Ciambelli:2021ujl} and including the present one, have all in common the end goal of formulating the correct gravitational framework to understand features of quantum gravity, such as entanglement entropy between subregions, non-factorizability, and the information conundrum. This is clearly a long road yet to pave, but we believe that the direction taken here is the correct one, for it completely disentangles classical features, like a metric or dynamics, from more fundamental properties, such as symmetries and their representations.

\paragraph{Acknowledgments}

We thank G. Barnich, L. Freidel and M. Klinger for discussions. We would also like to thank the participants of the first and second online Corners Meeting for stimulating discussions. LC acknowledges \'Ecole Polytechnique and Nordita for invitations to present this work. RGL thanks the Universit\'e Libre de Bruxelles for hospitality where a portion of this work was completed, as well as the opportunity to present the work at the Belgian joint seminars.
The research of LC was partially supported by a Marina Solvay 
Fellowship, by the ERC Advanced Grant ``High-Spin-Grav" and by 
FNRS-Belgium (convention FRFC PDR T.1025.14 and convention IISN 4.4503.15).
The work of RGL was supported by the U.S. Department of Energy under contract DE-SC0015655.

\appendix
\renewcommand{\theequation}{\thesection.\arabic{equation}}
\setcounter{equation}{0}

\section{Orbits Classification of $\mh$}\label{appA}

Although classical phase spaces are never mapped to the entirety of $\mh$ but only to its ideals $\mh_s$ and $\mh_w$, we consider the study of orbits in $\mh^*$ to be an important step toward the classification of representations and isotropy groups, so we report it in this appendix. 

As long as $C_3$ is non-vanishing and free to vary, the orbits are $6$-dimensional, because the KKS $2$-form can always be written as in \eqref{KKSdJdP}. This is confirmed by the Pfaffian being proportional to $C_3$ as in \eqref{pfaf}. So to study lower dimensional orbits, we have to start from the locus of points where
\beq\label{c30}
C_3= P_aJ^a{}_b\varepsilon^{bc}P_c=0.
\eeq
Let us study various solutions of this equation.
The rotational symmetry in the $2$-plane allows us to set e.g., $P_1=0$, which is the analogue of the particle rest frame of the Poincar\'e group analysis. Then, there are two ways to solve \eqref{c30}. 

The first is to set $P_0=0$. Eq. \eqref{dJdP} then can be inverted and gives rise to $2$ independent equations plus a constraint which, inserted back in the KKS $2$-form \eqref{KKS} give
\beqn\label{2do}
\Omega^{(\mh)}_{m, \ P_a=0}(\un\mu,\un\nu)
=\delta_{\un\mu}J^0{}_0\delta_{\un\nu}\log (J^1{}_0)-\delta_{\un\nu}J^0{}_0\delta_{\un\mu}\log (J^1{}_0).
\eeqn
Therefore, the general orbits at this point are $2$-dimensional. 

The second possibility is to set $J^0{}_1=0$. Then, supposing that $P_0\neq 0$, we can solve \eqref{dJdP}. The generic orbit is $4$-dimensional in this case. Introducing the combination
\beq
\delta_{\un\mu}\kappa\equiv {\delta_{\un\mu}P_1 J^1{}_0+\delta_{\un\mu}J^0{}_0P_0\over 2P_0^2},
\eeq
we obtain
\beqn\label{4do}
\Omega^{(\mh)}_{m, \ P_1=0, \ J^0{}_1=0}(\un\mu,\un\nu)
={\delta_{\un\mu}P_1\delta_{\un\nu}J^1{}_0+\delta_{\un\mu}(P_0)^2\delta_{\un\nu}\kappa-\delta_{\un\nu}P_1\delta_{\un\mu}J^1{}_0-\delta_{\un\nu}(P_0)^2\delta_{\un\mu}\kappa\over 2 P_0}.
\eeqn
From this, one can descend to $2$- and $0$-dimensional orbits by imposing further constraints. For instance, setting furthermore $P_0=0$ from the beginning gives exactly the $2$-dimensional orbits of \eqref{2do}. Note that one could have solved $C_3=0$ by setting $J^a{}_b=0$ and no restrictions on $P_a$. One then would have obtained a KKS $2$-form isomorphic to \eqref{4do}, as a consequence of rotational symmetry, as already pointed out.

We conclude by observing that $0$-dimensional orbits are obtained by giving at least $5$ constraints, which is a direct consequence of the fact that there are no a priori degenerate directions inside $\mh^*$, and so one has to go to very specific points to have maximal isotropy group.

\providecommand{\href}[2]{#2}\begingroup\raggedright\endgroup

\begin{thebibliography}{100}

\bibitem{Noether1918}
E.~Noether, ``Invariante variationsprobleme,''
  \href{http://dx.doi.org/doi:10.1080/00411457108231446}{{\em Nachrichten von
  der Gesellschaft der Wissenschaften\\ zu G\"ottingen,
  Mathematisch-Physikalische Klasse} {\bf 1918} (1918)  235--257}.

\bibitem{Regge:1974zd}
T.~Regge and C.~Teitelboim, ``{Role of Surface Integrals in the Hamiltonian
  Formulation of General Relativity},''
  \href{http://dx.doi.org/10.1016/0003-4916(74)90404-7}{{\em Ann. Phys.} {\bf
  88} (1974)  286}.

\bibitem{Donnelly:2016auv}
W.~Donnelly and L.~Freidel, ``{Local subsystems in gauge theory and gravity},''
  \href{http://dx.doi.org/10.1007/JHEP09(2016)102}{{\em JHEP} {\bf 09} (2016)
  102}, \href{http://arxiv.org/abs/1601.04744}{{\tt arXiv:1601.04744
  [hep-th]}}.

\bibitem{Speranza:2017gxd}
A.~J. Speranza, ``{Local phase space and edge modes for
  diffeomorphism-invariant theories},''
  \href{http://dx.doi.org/10.1007/JHEP02(2018)021}{{\em JHEP} {\bf 02} (2018)
  021}, \href{http://arxiv.org/abs/1706.05061}{{\tt arXiv:1706.05061
  [hep-th]}}.

\bibitem{Geiller:2017whh}
M.~Geiller, ``{Lorentz-diffeomorphism edge modes in 3d gravity},''
  \href{http://dx.doi.org/10.1007/JHEP02(2018)029}{{\em JHEP} {\bf 02} (2018)
  029}, \href{http://arxiv.org/abs/1712.05269}{{\tt arXiv:1712.05269 [gr-qc]}}.

\bibitem{Freidel:2020xyx}
L.~Freidel, M.~Geiller, and D.~Pranzetti, ``{Edge modes of gravity. Part I.
  Corner potentials and charges},''
  \href{http://dx.doi.org/10.1007/JHEP11(2020)026}{{\em JHEP} {\bf 11} (2020)
  026}, \href{http://arxiv.org/abs/2006.12527}{{\tt arXiv:2006.12527
  [hep-th]}}.

\bibitem{Freidel:2020svx}
L.~Freidel, M.~Geiller, and D.~Pranzetti, ``{Edge modes of gravity. Part II.
  Corner metric and Lorentz charges},''
  \href{http://dx.doi.org/10.1007/JHEP11(2020)027}{{\em JHEP} {\bf 11} (2020)
  027}, \href{http://arxiv.org/abs/2007.03563}{{\tt arXiv:2007.03563
  [hep-th]}}.

\bibitem{Freidel:2020ayo}
L.~Freidel, M.~Geiller, and D.~Pranzetti, ``{Edge modes of gravity. Part III.
  Corner simplicity constraints},''
  \href{http://dx.doi.org/10.1007/JHEP01(2021)100}{{\em JHEP} {\bf 01} (2021)
  100}, \href{http://arxiv.org/abs/2007.12635}{{\tt arXiv:2007.12635
  [hep-th]}}.

\bibitem{Donnelly:2020xgu}
W.~Donnelly, L.~Freidel, S.~F. Moosavian, and A.~J. Speranza, ``{Gravitational
  edge modes, coadjoint orbits, and hydrodynamics},''
  \href{http://dx.doi.org/10.1007/JHEP09(2021)008}{{\em JHEP} {\bf 09} (2021)
  008}, \href{http://arxiv.org/abs/2012.10367}{{\tt arXiv:2012.10367
  [hep-th]}}.

\bibitem{Ciambelli:2021vnn}
L.~Ciambelli and R.~G. Leigh, ``{Isolated surfaces and symmetries of
  gravity},'' \href{http://dx.doi.org/10.1103/PhysRevD.104.046005}{{\em Phys.
  Rev. D} {\bf 104} (2021) no.~4, 046005},
  \href{http://arxiv.org/abs/2104.07643}{{\tt arXiv:2104.07643 [hep-th]}}.

\bibitem{Freidel:2021cjp}
L.~Freidel, R.~Oliveri, D.~Pranzetti, and S.~Speziale, ``{Extended corner
  symmetry, charge bracket and Einstein's equations},''
  \href{http://dx.doi.org/10.1007/JHEP09(2021)083}{{\em JHEP} {\bf 09} (2021)
  083}, \href{http://arxiv.org/abs/2104.12881}{{\tt arXiv:2104.12881
  [hep-th]}}.

\bibitem{Ciambelli:2021nmv}
L.~Ciambelli, R.~G. Leigh, and P.-C. Pai, ``{Embeddings and Integrable Charges
  for Extended Corner Symmetry},''
  \href{http://dx.doi.org/10.1103/PhysRevLett.128.171302}{{\em Phys. Rev.
  Lett.} {\bf 128} (2022)  }, \href{http://arxiv.org/abs/2111.13181}{{\tt
  arXiv:2111.13181 [hep-th]}}.

\bibitem{Maldacena:1997re}
J.~M. Maldacena, ``{The Large N limit of superconformal field theories and
  supergravity},'' \href{http://dx.doi.org/10.1023/A:1026654312961}{{\em Adv.
  Theor. Math. Phys.} {\bf 2} (1998)  231--252},
  \href{http://arxiv.org/abs/hep-th/9711200}{{\tt arXiv:hep-th/9711200}}.

\bibitem{Witten:1998qj}
E.~Witten, ``{Anti-de Sitter space and holography},''
  \href{http://dx.doi.org/10.4310/ATMP.1998.v2.n2.a2}{{\em Adv. Theor. Math.
  Phys.} {\bf 2} (1998)  253--291},
  \href{http://arxiv.org/abs/hep-th/9802150}{{\tt arXiv:hep-th/9802150}}.

\bibitem{Brown:1986nw}
J.~D. Brown and M.~Henneaux, ``{Central Charges in the Canonical Realization of
  Asymptotic Symmetries: An Example from Three-Dimensional Gravity},''
  \href{http://dx.doi.org/10.1007/BF01211590}{{\em Commun. Math. Phys.} {\bf
  104} (1986)  207--226}.

\bibitem{Arcioni:2003xx}
G.~Arcioni and C.~Dappiaggi, ``{Exploring the holographic principle in
  asymptotically flat space-times via the BMS group},''
  \href{http://dx.doi.org/10.1016/j.nuclphysb.2003.09.051}{{\em Nucl. Phys. B}
  {\bf 674} (2003)  553--592}, \href{http://arxiv.org/abs/hep-th/0306142}{{\tt
  arXiv:hep-th/0306142}}.

\bibitem{Arcioni:2003td}
G.~Arcioni and C.~Dappiaggi, ``{Holography in asymptotically flat space-times
  and the BMS group},''
  \href{http://dx.doi.org/10.1088/0264-9381/21/23/022}{{\em Class. Quant.
  Grav.} {\bf 21} (2004)  5655},
  \href{http://arxiv.org/abs/hep-th/0312186}{{\tt arXiv:hep-th/0312186}}.

\bibitem{deBoer:2003vf}
J.~de~Boer and S.~N. Solodukhin, ``{A Holographic reduction of Minkowski
  space-time},'' \href{http://dx.doi.org/10.1016/S0550-3213(03)00494-2}{{\em
  Nucl. Phys. B} {\bf 665} (2003)  545--593},
  \href{http://arxiv.org/abs/hep-th/0303006}{{\tt arXiv:hep-th/0303006}}.

\bibitem{Dappiaggi:2005ci}
C.~Dappiaggi, V.~Moretti, and N.~Pinamonti, ``{Rigorous steps towards
  holography in asymptotically flat spacetimes},''
  \href{http://dx.doi.org/10.1142/S0129055X0600270X}{{\em Rev. Math. Phys.}
  {\bf 18} (2006)  349--416}, \href{http://arxiv.org/abs/gr-qc/0506069}{{\tt
  arXiv:gr-qc/0506069}}.

\bibitem{Barnich:2010eb}
G.~Barnich and C.~Troessaert, ``{Aspects of the BMS/CFT correspondence},''
  \href{http://dx.doi.org/10.1007/JHEP05(2010)062}{{\em JHEP} {\bf 05} (2010)
  062}, \href{http://arxiv.org/abs/1001.1541}{{\tt arXiv:1001.1541 [hep-th]}}.

\bibitem{Strominger:2013jfa}
A.~Strominger, ``{On BMS Invariance of Gravitational Scattering},''
  \href{http://dx.doi.org/10.1007/JHEP07(2014)152}{{\em JHEP} {\bf 07} (2014)
  152}, \href{http://arxiv.org/abs/1312.2229}{{\tt arXiv:1312.2229 [hep-th]}}.

\bibitem{Kapec:2016jld}
D.~Kapec, P.~Mitra, A.-M. Raclariu, and A.~Strominger, ``{2D Stress Tensor for
  4D Gravity},'' \href{http://dx.doi.org/10.1103/PhysRevLett.119.121601}{{\em
  Phys. Rev. Lett.} {\bf 119} (2017) no.~12, 121601},
  \href{http://arxiv.org/abs/1609.00282}{{\tt arXiv:1609.00282 [hep-th]}}.

\bibitem{Cheung:2016iub}
C.~Cheung, A.~de~la Fuente, and R.~Sundrum, ``{4D scattering amplitudes and
  asymptotic symmetries from 2D CFT},''
  \href{http://dx.doi.org/10.1007/JHEP01(2017)112}{{\em JHEP} {\bf 01} (2017)
  112}, \href{http://arxiv.org/abs/1609.00732}{{\tt arXiv:1609.00732
  [hep-th]}}.

\bibitem{Pasterski:2016qvg}
S.~Pasterski, S.-H. Shao, and A.~Strominger, ``{Flat Space Amplitudes and
  Conformal Symmetry of the Celestial Sphere},''
  \href{http://dx.doi.org/10.1103/PhysRevD.96.065026}{{\em Phys. Rev. D} {\bf
  96} (2017) no.~6, 065026}, \href{http://arxiv.org/abs/1701.00049}{{\tt
  arXiv:1701.00049 [hep-th]}}.

\bibitem{Pasterski:2017kqt}
S.~Pasterski and S.-H. Shao, ``{Conformal basis for flat space amplitudes},''
  \href{http://dx.doi.org/10.1103/PhysRevD.96.065022}{{\em Phys. Rev. D} {\bf
  96} (2017) no.~6, 065022}, \href{http://arxiv.org/abs/1705.01027}{{\tt
  arXiv:1705.01027 [hep-th]}}.

\bibitem{Donnay:2020guq}
L.~Donnay, S.~Pasterski, and A.~Puhm, ``{Asymptotic Symmetries and Celestial
  CFT},'' \href{http://dx.doi.org/10.1007/JHEP09(2020)176}{{\em JHEP} {\bf 09}
  (2020)  176}, \href{http://arxiv.org/abs/2005.08990}{{\tt arXiv:2005.08990
  [hep-th]}}.

\bibitem{Pate:2019lpp}
M.~Pate, A.-M. Raclariu, A.~Strominger, and E.~Y. Yuan, ``{Celestial operator
  products of gluons and gravitons},''
  \href{http://dx.doi.org/10.1142/S0129055X21400031}{{\em Rev. Math. Phys.}
  {\bf 33} (2021) no.~09, 2140003}, \href{http://arxiv.org/abs/1910.07424}{{\tt
  arXiv:1910.07424 [hep-th]}}.

\bibitem{Fotopoulos:2019vac}
A.~Fotopoulos, S.~Stieberger, T.~R. Taylor, and B.~Zhu, ``{Extended BMS Algebra
  of Celestial CFT},'' \href{http://dx.doi.org/10.1007/JHEP03(2020)130}{{\em
  JHEP} {\bf 03} (2020)  130}, \href{http://arxiv.org/abs/1912.10973}{{\tt
  arXiv:1912.10973 [hep-th]}}.

\bibitem{Donnay:2021wrk}
L.~Donnay and R.~Ruzziconi, ``{BMS flux algebra in celestial holography},''
  \href{http://dx.doi.org/10.1007/JHEP11(2021)040}{{\em JHEP} {\bf 11} (2021)
  040}, \href{http://arxiv.org/abs/2108.11969}{{\tt arXiv:2108.11969
  [hep-th]}}.

\bibitem{Pasterski:2021raf}
S.~Pasterski, M.~Pate, and A.-M. Raclariu, ``{Celestial Holography},'' in {\em
  {2022 Snowmass Summer Study}}.
\newblock 11, 2021.
\newblock \href{http://arxiv.org/abs/2111.11392}{{\tt arXiv:2111.11392
  [hep-th]}}.

\bibitem{Bagchi:2010zz}
A.~Bagchi, ``{Correspondence between Asymptotically Flat Spacetimes and
  Nonrelativistic Conformal Field Theories},''
  \href{http://dx.doi.org/10.1103/PhysRevLett.105.171601}{{\em Phys. Rev.
  Lett.} {\bf 105} (2010)  171601}, \href{http://arxiv.org/abs/1006.3354}{{\tt
  arXiv:1006.3354 [hep-th]}}.

\bibitem{Hartong:2015xda}
J.~Hartong, ``{Gauging the Carroll Algebra and Ultra-Relativistic Gravity},''
  \href{http://dx.doi.org/10.1007/JHEP08(2015)069}{{\em JHEP} {\bf 08} (2015)
  069}, \href{http://arxiv.org/abs/1505.05011}{{\tt arXiv:1505.05011
  [hep-th]}}.

\bibitem{Ciambelli:2018wre}
L.~Ciambelli, C.~Marteau, A.~C. Petkou, P.~M. Petropoulos, and K.~Siampos,
  ``{Flat holography and Carrollian fluids},''
  \href{http://dx.doi.org/10.1007/JHEP07(2018)165}{{\em JHEP} {\bf 07} (2018)
  165}, \href{http://arxiv.org/abs/1802.06809}{{\tt arXiv:1802.06809
  [hep-th]}}.

\bibitem{Ciambelli:2018ojf}
L.~Ciambelli and C.~Marteau, ``{Carrollian conservation laws and Ricci-flat
  gravity},'' \href{http://dx.doi.org/10.1088/1361-6382/ab0d37}{{\em Class.
  Quant. Grav.} {\bf 36} (2019) no.~8, 085004},
  \href{http://arxiv.org/abs/1810.11037}{{\tt arXiv:1810.11037 [hep-th]}}.

\bibitem{Campoleoni:2018ltl}
A.~Campoleoni, L.~Ciambelli, C.~Marteau, P.~M. Petropoulos, and K.~Siampos,
  ``{Two-dimensional fluids and their holographic duals},''
  \href{http://dx.doi.org/10.1016/j.nuclphysb.2019.114692}{{\em Nucl. Phys. B}
  {\bf 946} (2019)  114692}, \href{http://arxiv.org/abs/1812.04019}{{\tt
  arXiv:1812.04019 [hep-th]}}.

\bibitem{Figueroa-OFarrill:2021sxz}
J.~Figueroa-O'Farrill, E.~Have, S.~Prohazka, and J.~Salzer, ``{Carrollian and
  celestial spaces at infinity},'' \href{http://arxiv.org/abs/2112.03319}{{\tt
  arXiv:2112.03319 [hep-th]}}.

\bibitem{Donnay:2022aba}
L.~Donnay, A.~Fiorucci, Y.~Herfray, and R.~Ruzziconi, ``{A Carrollian
  Perspective on Celestial Holography},''
  \href{http://arxiv.org/abs/2202.04702}{{\tt arXiv:2202.04702 [hep-th]}}.

\bibitem{Hopfmuller:2016scf}
F.~Hopfm\"uller and L.~Freidel, ``{Gravity Degrees of Freedom on a Null
  Surface},'' \href{http://dx.doi.org/10.1103/PhysRevD.95.104006}{{\em Phys.
  Rev. D} {\bf 95} (2017) no.~10, 104006},
  \href{http://arxiv.org/abs/1611.03096}{{\tt arXiv:1611.03096 [gr-qc]}}.

\bibitem{Hopfmuller:2018fni}
F.~Hopfm\"uller and L.~Freidel, ``{Null Conservation Laws for Gravity},''
  \href{http://dx.doi.org/10.1103/PhysRevD.97.124029}{{\em Phys. Rev. D} {\bf
  97} (2018) no.~12, 124029}, \href{http://arxiv.org/abs/1802.06135}{{\tt
  arXiv:1802.06135 [gr-qc]}}.

\bibitem{Chandrasekaran:2018aop}
V.~Chandrasekaran, E.~E. Flanagan, and K.~Prabhu, ``{Symmetries and charges of
  general relativity at null boundaries},''
  \href{http://dx.doi.org/10.1007/JHEP11(2018)125}{{\em JHEP} {\bf 11} (2018)
  125}, \href{http://arxiv.org/abs/1807.11499}{{\tt arXiv:1807.11499
  [hep-th]}}.

\bibitem{Ciambelli:2019lap}
L.~Ciambelli, R.~G. Leigh, C.~Marteau, and P.~M. Petropoulos, ``{Carroll
  Structures, Null Geometry and Conformal Isometries},''
  \href{http://dx.doi.org/10.1103/PhysRevD.100.046010}{{\em Phys. Rev. D} {\bf
  100} (2019) no.~4, 046010}, \href{http://arxiv.org/abs/1905.02221}{{\tt
  arXiv:1905.02221 [hep-th]}}.

\bibitem{Speranza:2019hkr}
A.~J. Speranza, ``{Geometrical tools for embedding fields, submanifolds, and
  foliations},'' \href{http://arxiv.org/abs/1904.08012}{{\tt arXiv:1904.08012
  [gr-qc]}}.

\bibitem{Adami:2021nnf}
H.~Adami, D.~Grumiller, M.~M. Sheikh-Jabbari, V.~Taghiloo, H.~Yavartanoo, and
  C.~Zwikel, ``{Null boundary phase space: slicings, news \& memory},''
  \href{http://dx.doi.org/10.1007/JHEP11(2021)155}{{\em JHEP} {\bf 11} (2021)
  155}, \href{http://arxiv.org/abs/2110.04218}{{\tt arXiv:2110.04218
  [hep-th]}}.

\bibitem{Chandrasekaran:2021hxc}
V.~Chandrasekaran, E.~E. Flanagan, I.~Shehzad, and A.~J. Speranza,
  ``{Brown-York charges at null boundaries},''
  \href{http://dx.doi.org/10.1007/JHEP01(2022)029}{{\em JHEP} {\bf 01} (2022)
  029}, \href{http://arxiv.org/abs/2109.11567}{{\tt arXiv:2109.11567
  [hep-th]}}.

\bibitem{Donnay:2015abr}
L.~Donnay, G.~Giribet, H.~A. Gonzalez, and M.~Pino, ``{Supertranslations and
  Superrotations at the Black Hole Horizon},''
  \href{http://dx.doi.org/10.1103/PhysRevLett.116.091101}{{\em Phys. Rev.
  Lett.} {\bf 116} (2016) no.~9, 091101},
  \href{http://arxiv.org/abs/1511.08687}{{\tt arXiv:1511.08687 [hep-th]}}.

\bibitem{Donnay:2016ejv}
L.~Donnay, G.~Giribet, H.~A. Gonz\'alez, and M.~Pino, ``{Extended Symmetries at
  the Black Hole Horizon},''
  \href{http://dx.doi.org/10.1007/JHEP09(2016)100}{{\em JHEP} {\bf 09} (2016)
  100}, \href{http://arxiv.org/abs/1607.05703}{{\tt arXiv:1607.05703
  [hep-th]}}.

\bibitem{Donnay:2019jiz}
L.~Donnay and C.~Marteau, ``{Carrollian Physics at the Black Hole Horizon},''
  \href{http://dx.doi.org/10.1088/1361-6382/ab2fd5}{{\em Class. Quant. Grav.}
  {\bf 36} (2019) no.~16, 165002}, \href{http://arxiv.org/abs/1903.09654}{{\tt
  arXiv:1903.09654 [hep-th]}}.

\bibitem{Grumiller:2019fmp}
D.~Grumiller, A.~P\'erez, M.~M. Sheikh-Jabbari, R.~Troncoso, and C.~Zwikel,
  ``{Spacetime structure near generic horizons and soft hair},''
  \href{http://dx.doi.org/10.1103/PhysRevLett.124.041601}{{\em Phys. Rev.
  Lett.} {\bf 124} (2020) no.~4, 041601},
  \href{http://arxiv.org/abs/1908.09833}{{\tt arXiv:1908.09833 [hep-th]}}.

\bibitem{Carlip:2019dbu}
S.~Carlip, ``{Near-horizon Bondi-Metzner-Sachs symmetry, dimensional reduction,
  and black hole entropy},''
  \href{http://dx.doi.org/10.1103/PhysRevD.101.046002}{{\em Phys. Rev. D} {\bf
  101} (2020) no.~4, 046002}, \href{http://arxiv.org/abs/1910.01762}{{\tt
  arXiv:1910.01762 [hep-th]}}.

\bibitem{Sachs:1962wk}
R.~Sachs, ``{Gravitational Waves in General Relativity. 8. Waves in
  asymptotically flat space-times},''
  \href{http://dx.doi.org/10.1098/rspa.1962.0206}{{\em Proc. Roy. Soc. Lond. A}
  {\bf 270} (1962)  103--126}.

\bibitem{doi:10.1098/rspa.1962.0161}
H.~Bondi, M.~G.~J. Van~der Burg, and A.~W.~K. Metzner, ``{Gravitational waves
  in general relativity, VII. Waves from axi-symmetric isolated system},''
  \href{http://dx.doi.org/10.1098/rspa.1962.0161}{{\em Proc. Roy. Soc. Lond. A}
  {\bf 269} (1962) no.~1336, 21--52}.

\bibitem{doi:10.1098/rspa.1962.0206}
R.~Sachs and H.~Bondi, ``{Gravitational Waves in General Relativity VIII. Waves
  in Asymptotically Flat Space-time},''
  \href{http://dx.doi.org/10.1098/rspa.1962.0206}{{\em Proceedings of the Royal
  Society of London. Series A. Mathematical and Physical\\ Sciences} {\bf 270}
  (1962) no.~1340, 103--126}.

\bibitem{Campiglia:2014yka}
M.~Campiglia and A.~Laddha, ``{Asymptotic symmetries and subleading soft
  graviton theorem},'' \href{http://dx.doi.org/10.1103/PhysRevD.90.124028}{{\em
  Phys. Rev. D} {\bf 90} (2014) no.~12, 124028},
  \href{http://arxiv.org/abs/1408.2228}{{\tt arXiv:1408.2228 [hep-th]}}.

\bibitem{Compere:2018ylh}
G.~Comp\`ere, A.~Fiorucci, and R.~Ruzziconi, ``{Superboost transitions,
  refraction memory and super-Lorentz charge algebra},''
  \href{http://dx.doi.org/10.1007/JHEP11(2018)200}{{\em JHEP} {\bf 11} (2018)
  200}, \href{http://arxiv.org/abs/1810.00377}{{\tt arXiv:1810.00377
  [hep-th]}}. [Erratum: JHEP 04, 172 (2020)].

\bibitem{Campiglia:2020qvc}
M.~Campiglia and J.~Peraza, ``{Generalized BMS charge algebra},''
  \href{http://dx.doi.org/10.1103/PhysRevD.101.104039}{{\em Phys. Rev. D} {\bf
  101} (2020) no.~10, 104039}, \href{http://arxiv.org/abs/2002.06691}{{\tt
  arXiv:2002.06691 [gr-qc]}}.

\bibitem{Flanagan:2019vbl}
E.~E. Flanagan, K.~Prabhu, and I.~Shehzad, ``{Extensions of the asymptotic
  symmetry algebra of general relativity},''
  \href{http://dx.doi.org/10.1007/JHEP01(2020)002}{{\em JHEP} {\bf 01} (2020)
  002}, \href{http://arxiv.org/abs/1910.04557}{{\tt arXiv:1910.04557 [gr-qc]}}.

\bibitem{Freidel:2021fxf}
L.~Freidel, R.~Oliveri, D.~Pranzetti, and S.~Speziale, ``{The Weyl BMS group
  and Einstein's equations},''
  \href{http://dx.doi.org/10.1007/JHEP07(2021)170}{{\em JHEP} {\bf 07} (2021)
  170}, \href{http://arxiv.org/abs/2104.05793}{{\tt arXiv:2104.05793
  [hep-th]}}.

\bibitem{Troessaert:2013fma}
C.~Troessaert, ``{Enhanced asymptotic symmetry algebra of $AdS_{3}$},''
  \href{http://dx.doi.org/10.1007/JHEP08(2013)044}{{\em JHEP} {\bf 08} (2013)
  044}, \href{http://arxiv.org/abs/1303.3296}{{\tt arXiv:1303.3296 [hep-th]}}.

\bibitem{Grumiller:2016pqb}
D.~Grumiller and M.~Riegler, ``{Most general AdS$_{3}$ boundary conditions},''
  \href{http://dx.doi.org/10.1007/JHEP10(2016)023}{{\em JHEP} {\bf 10} (2016)
  023}, \href{http://arxiv.org/abs/1608.01308}{{\tt arXiv:1608.01308
  [hep-th]}}.

\bibitem{Ciambelli:2019bzz}
L.~Ciambelli and R.~G. Leigh, ``{Weyl Connections and their Role in
  Holography},'' \href{http://dx.doi.org/10.1103/PhysRevD.101.086020}{{\em
  Phys. Rev. D} {\bf 101} (2020) no.~8, 086020},
  \href{http://arxiv.org/abs/1905.04339}{{\tt arXiv:1905.04339 [hep-th]}}.

\bibitem{Compere:2019bua}
G.~Comp\`ere, A.~Fiorucci, and R.~Ruzziconi, ``{The $\Lambda$-BMS$_4$ group of
  dS$_4$ and new boundary conditions for AdS$_4$},''
  \href{http://dx.doi.org/10.1088/1361-6382/ab3d4b}{{\em Class. Quant. Grav.}
  {\bf 36} (2019) no.~19, 195017}, \href{http://arxiv.org/abs/1905.00971}{{\tt
  arXiv:1905.00971 [gr-qc]}}. [Erratum: Class.Quant.Grav. 38, 229501 (2021)].

\bibitem{Alessio:2020ioh}
F.~Alessio, G.~Barnich, L.~Ciambelli, P.~Mao, and R.~Ruzziconi, ``{Weyl charges
  in asymptotically locally AdS$_3$ spacetimes},''
  \href{http://dx.doi.org/10.1103/PhysRevD.103.046003}{{\em Phys. Rev. D} {\bf
  103} (2021) no.~4, 046003}, \href{http://arxiv.org/abs/2010.15452}{{\tt
  arXiv:2010.15452 [hep-th]}}.

\bibitem{Fiorucci:2020xto}
A.~Fiorucci and R.~Ruzziconi, ``{Charge algebra in Al(A)dS$_{n}$ spacetimes},''
  \href{http://dx.doi.org/10.1007/JHEP05(2021)210}{{\em JHEP} {\bf 05} (2021)
  210}, \href{http://arxiv.org/abs/2011.02002}{{\tt arXiv:2011.02002
  [hep-th]}}.

\bibitem{Ciambelli:2020eba}
L.~Ciambelli, C.~Marteau, P.~M. Petropoulos, and R.~Ruzziconi, ``{Gauges in
  Three-Dimensional Gravity and Holographic Fluids},''
  \href{http://dx.doi.org/10.1007/JHEP11(2020)092}{{\em JHEP} {\bf 11} (2020)
  092}, \href{http://arxiv.org/abs/2006.10082}{{\tt arXiv:2006.10082
  [hep-th]}}.

\bibitem{Geiller:2022vto}
M.~Geiller and C.~Zwikel, ``{The partial Bondi gauge: Further enlarging the
  asymptotic structure of gravity},''
  \href{http://arxiv.org/abs/2205.11401}{{\tt arXiv:2205.11401 [hep-th]}}.

\bibitem{Witten:1987ty}
E.~Witten, ``{Coadjoint Orbits of the Virasoro Group},''
  \href{http://dx.doi.org/10.1007/BF01218287}{{\em Commun. Math. Phys.} {\bf
  114} (1988)  1}.

\bibitem{Balog:1997zz}
J.~Balog, L.~Feher, and L.~Palla, ``{Coadjoint orbits of the Virasoro algebra
  and the global Liouville equation},''
  \href{http://dx.doi.org/10.1142/S0217751X98000147}{{\em Int. J. Mod. Phys. A}
  {\bf 13} (1998)  315--362}, \href{http://arxiv.org/abs/hep-th/9703045}{{\tt
  arXiv:hep-th/9703045}}.

\bibitem{Barnich:2014zoa}
G.~Barnich and B.~Oblak, ``{Holographic positive energy theorems in
  three-dimensional gravity},''
  \href{http://dx.doi.org/10.1088/0264-9381/31/15/152001}{{\em Class. Quant.
  Grav.} {\bf 31} (2014)  152001}, \href{http://arxiv.org/abs/1403.3835}{{\tt
  arXiv:1403.3835 [hep-th]}}.

\bibitem{Barnich:2014kra}
G.~Barnich and B.~Oblak, ``{Notes on the BMS group in three dimensions: I.
  Induced representations},''
  \href{http://dx.doi.org/10.1007/JHEP06(2014)129}{{\em JHEP} {\bf 06} (2014)
  129}, \href{http://arxiv.org/abs/1403.5803}{{\tt arXiv:1403.5803 [hep-th]}}.

\bibitem{Duval:2014lpa}
C.~Duval, G.~W. Gibbons, and P.~A. Horvathy, ``{Conformal Carroll groups},''
  \href{http://dx.doi.org/10.1088/1751-8113/47/33/335204}{{\em J. Phys. A} {\bf
  47} (2014) no.~33, 335204}, \href{http://arxiv.org/abs/1403.4213}{{\tt
  arXiv:1403.4213 [hep-th]}}.

\bibitem{Barnich:2015uva}
G.~Barnich and B.~Oblak, ``{Notes on the BMS group in three dimensions: II.
  Coadjoint representation},''
  \href{http://dx.doi.org/10.1007/JHEP03(2015)033}{{\em JHEP} {\bf 03} (2015)
  033}, \href{http://arxiv.org/abs/1502.00010}{{\tt arXiv:1502.00010
  [hep-th]}}.

\bibitem{Barnich:2021dta}
G.~Barnich and R.~Ruzziconi, ``{Coadjoint representation of the BMS group on
  celestial Riemann surfaces},''
  \href{http://dx.doi.org/10.1007/JHEP06(2021)079}{{\em JHEP} {\bf 06} (2021)
  079}, \href{http://arxiv.org/abs/2103.11253}{{\tt arXiv:2103.11253 [gr-qc]}}.

\bibitem{Lahlali:2021nrf}
I.~A. Lahlali, N.~Boulanger, and A.~Campoleoni, ``{Coadjoint Orbits of the
  Poincar\'e Group for Discrete-Spin Particles in Any Dimension},''
  \href{http://dx.doi.org/10.3390/sym13091749}{{\em Symmetry} {\bf 13} (2021)
  no.~9, 1749}.

\bibitem{Marsot:2021tvq}
L.~Marsot, ``{Planar Carrollean dynamics, and the Carroll quantum equation},''
  \href{http://dx.doi.org/10.1016/j.geomphys.2022.104574}{{\em J. Geom. Phys.}
  {\bf 179} (2022)  104574}, \href{http://arxiv.org/abs/2110.08489}{{\tt
  arXiv:2110.08489 [math-ph]}}.

\bibitem{Bergshoeff:2022eog}
E.~Bergshoeff, J.~Figueroa-O'Farrill, and J.~Gomis, ``{A non-lorentzian
  primer},'' \href{http://arxiv.org/abs/2206.12177}{{\tt arXiv:2206.12177
  [hep-th]}}.

\bibitem{Riello:2022din}
A.~Riello and M.~Schiavina, ``{Hamiltonian gauge theory with corners:
  constraint reduction and flux superselection},''
  \href{http://arxiv.org/abs/2207.00568}{{\tt arXiv:2207.00568 [math-ph]}}.

\bibitem{Oblak:2016eij}
B.~Oblak, \href{http://dx.doi.org/10.1007/978-3-319-61878-4}{{\em {BMS
  Particles in Three Dimensions}}}.
\newblock PhD thesis, U. Brussels, Brussels U., 2016.
\newblock \href{http://arxiv.org/abs/1610.08526}{{\tt arXiv:1610.08526
  [hep-th]}}.

\bibitem{Kirillov_1962}
A.~A. Kirillov, ``{Unitary Representations of Nilpotent Lie Groups},''
  \href{http://dx.doi.org/10.1070/rm1962v017n04abeh004118}{{\em Russian
  Mathematical Surveys} {\bf 17} (1962) no.~4, 53--104}.

\bibitem{Kirillov1976ElementsOT}
A.~A. Kirillov, \href{http://dx.doi.org/doi.org/10.1007/978-3-642-66243-0}{{\em
  {Elements of the Theory of Representations}}}.
\newblock Springer Berlin Heidelberg, 1976.

\bibitem{Kirillov_Merits}
A.~A. Kirillov, ``{Merits and Demerits of the Orbit Method},''
  \href{http://dx.doi.org/10.1090/S0273-0979-99-00849-6}{{\em Bull. Amer. Math.
  Soc.} {\bf 36} (1999)  433--488}.

\bibitem{Kirillov1990}
A.~A. Kirillov, {\em {Geometric Quantization}}, vol.~4 of {\em Encyclopaedia of
  Mathematical Sciences},
  \href{http://dx.doi.org/10.1007/978-3-662-06793-2_2}{pp.~139--176}.
\newblock Springer Berlin Heidelberg, 1990.

\bibitem{kirillov2004lectures}
A.~A. Kirillov, \href{http://dx.doi.org/10.1090/gsm/064}{{\em {Lectures on the
  Orbit Method}}}, vol.~64 of {\em Graduate Studies in Mathematics}.
\newblock American Mathematical Society, 2004.

\bibitem{Kostant_2006}
B.~Kostant, \href{http://dx.doi.org/10.1007/BFb0079068}{``Quantization and
  unitary representations,''} in {\em Lectures in Modern Analysis and
  Applications III}, vol.~170 of {\em Lecture Notes in Mathematics},
  pp.~87--208.
\newblock Springer Berlin Heidelberg, 2006.

\bibitem{ginz}
V.~A. Ginzburg, ``Method of orbits in the representation theory of complex lie
  groups,'' \href{http://dx.doi.org/10.1007/BF01082375}{{\em Functional
  Analysis and Its Applications} {\bf 15} (1981) no.~1, 18--28}.

\bibitem{Duistermaat:1982aa}
J.~J. Duistermaat and G.~J. Heckman, ``On the variation in the cohomology of
  the symplectic form of the reduced phase space,''
  \href{http://dx.doi.org/10.1007/BF01399506}{{\em Inventiones mathematicae}
  {\bf 69} (1982) no.~2, 259--268}.

\bibitem{ALEKSEEV1989719}
A.~Alekseev and S.~Shatashvili, ``Path integral quantization of the coadjoint
  orbits of the virasoro group and 2-d gravity,''
  \href{http://dx.doi.org/10.1016/0550-3213(89)90130-2}{{\em Nucl. Phys. B}
  {\bf 323} (1989) no.~3, 719--733}.

\bibitem{wildberger_1990}
N.~J. Wildberger, ``On a relationship between adjoint orbits and conjugacy
  classes of a lie group,''
  \href{http://dx.doi.org/10.4153/CMB-1990-048-4}{{\em Can. Math. Bull.} {\bf
  33} (1990) no.~3, 297--304}.

\bibitem{Brylinski_1994}
R.~Brylinski and B.~Kostant, ``Minimal representations, geometric quantization,
  and unitarity.,'' \href{http://dx.doi.org/10.1073/pnas.91.13.6026}{{\em Proc.
  Nat. Acad. Sci.} {\bf 91} (1994) no.~13, 6026--6029}.

\bibitem{souriau1970structure}
J.~Souriau, {\em Structure des syst{\`e}mes dynamiques}.
\newblock Collection Dunod Universit{\'e}. Dunod, 1970.

\bibitem{Kostant2009}
B.~Kostant, {\em {Orbits, Symplectic Structures and Representation Theory}},
  vol.~1 of {\em Collected Papers},
  \href{http://dx.doi.org/10.1007/b94535}{p.~482}.
\newblock Springer New York, 2009.

\bibitem{Atiyah:1957}
M.~F. Atiyah, ``{Complex Analytic Connections in Fibre Bundles},''
  \href{http://dx.doi.org/10.2307/1992969}{{\em Trans. Amer. Math. Soc.} {\bf
  85} (1957) no.~1, 181--207}.

\bibitem{Atiyah:1979iu}
M.~Atiyah, {\em {Geometry of Yang-Mills Fields}}.
\newblock Sc. Norm. Sup. (1979) Lezioni Fermiane, 1979.

\bibitem{mackenzie_1987}
K.~C.~H. Mackenzie, \href{http://dx.doi.org/10.1017/CBO9780511661839}{{\em {Lie
  Groupoids and Lie Algebroids in Differential Geometry}}}.
\newblock London Mathematical Society Lecture Note Series. Cambridge University
  Press, 1987.

\bibitem{mackenzie_2005}
K.~C.~H. Mackenzie, \href{http://dx.doi.org/10.1017/CBO9781107325883}{{\em
  {General Theory of Lie Groupoids and Lie Algebroids}}}.
\newblock London Mathematical Society Lecture Note Series. Cambridge University
  Press, 2005.

\bibitem{Lazzarini_2012}
S.~Lazzarini and T.~Masson, ``Connections on lie algebroids and on
  derivation-based noncommutative geometry,''
  \href{http://dx.doi.org/10.1016/j.geomphys.2011.11.002}{{\em J. Geom. Phys.}
  {\bf 62} (2012) no.~2, 387 -- 402}.

\bibitem{Fournel:2012uv}
C.~Fournel, S.~Lazzarini, and T.~Masson, ``{Formulation of gauge theories on
  transitive Lie algebroids},''
  \href{http://dx.doi.org/10.1016/j.geomphys.2012.11.005}{{\em J. Geom. Phys.}
  {\bf 64} (2013)  174--191}, \href{http://arxiv.org/abs/1205.6725}{{\tt
  arXiv:1205.6725 [math-ph]}}.

\bibitem{Jordan:2014uza}
F.~Jordan, S.~Lazzarini, and T.~Masson, {\em {Gauge field theories: various
  mathematical approaches}}.
\newblock Copernicus Center Press Krakow, 2014.
\newblock \href{http://arxiv.org/abs/1404.4604}{{\tt 1404.4604 [math-ph]}}.

\bibitem{Carow-Watamura:2016lob}
U.~Carow-Watamura, M.~A. Heller, N.~Ikeda, T.~Kaneko, and S.~Watamura,
  ``{Off-Shell Covariantization of Algebroid Gauge Theories},''
  \href{http://dx.doi.org/10.1093/ptep/ptx100}{{\em PTEP} {\bf 2017} (2017)
  no.~8, 083B01}, \href{http://arxiv.org/abs/1612.02612}{{\tt arXiv:1612.02612
  [hep-th]}}.

\bibitem{Kotov:2016lpx}
A.~Kotov and T.~Strobl, ``{Lie algebroids, gauge theories, and compatible
  geometrical structures},''
  \href{http://dx.doi.org/10.1142/S0129055X19500156}{{\em Rev. Math. Phys.}
  {\bf 31} (2018) no.~04, 1950015}, \href{http://arxiv.org/abs/1603.04490}{{\tt
  arXiv:1603.04490 [math.DG]}}.

\bibitem{Attard:2019pvw}
J.~Attard, J.~Fran\c{c}ois, S.~Lazzarini, and T.~Masson, ``{Cartan Connections
  and Atiyah Lie Algebroids},''
  \href{http://dx.doi.org/10.1016/j.geomphys.2019.103541}{{\em J. Geom. Phys.}
  {\bf 148} (2020)  103541}, \href{http://arxiv.org/abs/1904.04915}{{\tt
  arXiv:1904.04915 [math-ph]}}.

\bibitem{Ciambelli:2021ujl}
L.~Ciambelli and R.~G. Leigh, ``{Lie algebroids and the geometry of off-shell
  BRST},'' \href{http://dx.doi.org/10.1016/j.nuclphysb.2021.115553}{{\em Nucl.
  Phys. B} {\bf 972} (2021)  115553},
  \href{http://arxiv.org/abs/2101.03974}{{\tt arXiv:2101.03974 [hep-th]}}.

\bibitem{Freidel:2021dxw}
L.~Freidel, ``{A canonical bracket for open gravitational system},''
  \href{http://arxiv.org/abs/2111.14747}{{\tt arXiv:2111.14747 [hep-th]}}.

\bibitem{Lee:1990nz}
J.~Lee and R.~M. Wald, ``{Local symmetries and constraints},''
  \href{http://dx.doi.org/10.1063/1.528801}{{\em J. Math. Phys.} {\bf 31}
  (1990)  725--743}.

\bibitem{Wald:1993nt}
R.~M. Wald, ``{Black hole entropy is the Noether charge},''
  \href{http://dx.doi.org/10.1103/PhysRevD.48.R3427}{{\em Phys. Rev. D} {\bf
  48} (1993) no.~8, 3427--3431}, \href{http://arxiv.org/abs/gr-qc/9307038}{{\tt
  arXiv:gr-qc/9307038}}.

\bibitem{Iyer:1994ys}
V.~Iyer and R.~M. Wald, ``{Some properties of Noether charge and a proposal for
  dynamical black hole entropy},''
  \href{http://dx.doi.org/10.1103/PhysRevD.50.846}{{\em Phys. Rev. D} {\bf 50}
  (1994)  846--864}, \href{http://arxiv.org/abs/gr-qc/9403028}{{\tt
  arXiv:gr-qc/9403028}}.

\bibitem{Wald:1999wa}
R.~M. Wald and A.~Zoupas, ``{A General definition of 'conserved quantities' in
  general relativity and other theories of gravity},''
  \href{http://dx.doi.org/10.1103/PhysRevD.61.084027}{{\em Phys. Rev. D} {\bf
  61} (2000)  084027}, \href{http://arxiv.org/abs/gr-qc/9911095}{{\tt
  arXiv:gr-qc/9911095}}.

\bibitem{Compere:2019qed}
G.~Comp\`ere, \href{http://dx.doi.org/10.1007/978-3-030-04260-8}{{\em {Advanced
  Lectures on General Relativity}}}, vol.~952 of {\em Lecture Notes in
  Physics}.
\newblock Springer, 2019.

\bibitem{Harlow:2019yfa}
D.~Harlow and J.-Q. Wu, ``{Covariant phase space with boundaries},''
  \href{http://dx.doi.org/10.1007/JHEP10(2020)146}{{\em JHEP} {\bf 10} (2020)
  146}, \href{http://arxiv.org/abs/1906.08616}{{\tt arXiv:1906.08616
  [hep-th]}}.

\bibitem{Barnich:2001jy}
G.~Barnich and F.~Brandt, ``{Covariant theory of asymptotic symmetries,
  conservation laws and central charges},''
  \href{http://dx.doi.org/10.1016/S0550-3213(02)00251-1}{{\em Nucl. Phys. B}
  {\bf 633} (2002)  3--82}, \href{http://arxiv.org/abs/hep-th/0111246}{{\tt
  arXiv:hep-th/0111246}}.

\bibitem{Freidel:2015gpa}
L.~Freidel and A.~Perez, ``{Quantum gravity at the corner},''
  \href{http://dx.doi.org/10.3390/universe4100107}{{\em Universe} {\bf 4}
  (2018) no.~10, 107}, \href{http://arxiv.org/abs/1507.02573}{{\tt
  arXiv:1507.02573 [gr-qc]}}.

\bibitem{Speranza:2022lxr}
A.~J. Speranza, ``{Ambiguity resolution for integrable gravitational
  charges},'' \href{http://dx.doi.org/10.1007/JHEP07(2022)029}{{\em JHEP} {\bf
  07} (2022)  029}, \href{http://arxiv.org/abs/2202.00133}{{\tt
  arXiv:2202.00133 [hep-th]}}.

\bibitem{Carrozza:2022xut}
S.~Carrozza, S.~Eccles, and P.~A. Hoehn, ``{Edge modes as dynamical frames:
  charges from post-selection in generally covariant theories},''
  \href{http://arxiv.org/abs/2205.00913}{{\tt arXiv:2205.00913 [hep-th]}}.

\bibitem{Kabel:2022efn}
V.~Kabel and W.~Wieland, ``{Metriplectic geometry for gravitational
  subsystems},'' \href{http://arxiv.org/abs/2206.00029}{{\tt arXiv:2206.00029
  [gr-qc]}}.

\bibitem{Goeller:2022rsx}
C.~Goeller, P.~A. Hoehn, and J.~Kirklin, ``{Diffeomorphism-invariant
  observables and dynamical frames in gravity: reconciling bulk locality with
  general covariance},'' \href{http://arxiv.org/abs/2206.01193}{{\tt
  arXiv:2206.01193 [hep-th]}}.

\end{thebibliography}
\end{document}